\documentclass[runningheads,citeauthoryear]{apinv}
\usepackage{epsfig,cite,graphics}
\usepackage{marvosym} %For astronomical symbols % % %
\usepackage[utf8]{inputenc}
\usepackage{subcaption}

\begin{document}

\title{Structure of Open Clusters - \textit{Gaia} DR2 and its limitations}
\titlerunning{Structure of Open Clusters - \textit{Gaia}}
\author{Martin Piecka\inst{1}, Ernst Paunzen\inst{1}}
\authorrunning{M.Piecka, E.Paunzen}
\tocauthor{Martin Piecka, Ernst Paunzen} 
% Command tocautor{} is used by the Latex to give author names 
% to the Contents of the volume (automatically generated)
\institute{Department of Theoretical Physics and Astrophysics, Masaryk University, Kotl\'a\v{r}sk\'a 2, CZ-611\,37 Brno, Czech Republic
	\newline
	\email{408988@mail.muni.cz}    }
\papertype{Submitted on xx.xx.xxxx; Accepted on xx.xx.xxxx}	
% Papertype can be "Research report", "Review", "Invited lecture", "Conference talk", 
% "Conference poster", "Lecture at scientific seminar", "Summary of dissertation",  etc.
\maketitle

\begin{abstract}
Very precise observational data are needed for studying the stellar cluster parameters (distance, reddening, age, metallicity) and cluster internal kinematics. In turn, these give us an insight into the properties of our Galaxy, for example, by giving us the ability to trace Galactic spiral structure, star formation rates and metallicity gradients. We investigated the available \textit{Gaia}~DR2 catalogue of 1229 open clusters and studied cluster distances, sizes and membership distributions in the 3D space. An appropriate analysis of the parallax-to-distance transformation problem is presented in the context of getting distances toward open clusters and estimating their sizes. Based on our investigation of the \textit{Gaia}~DR2 data we argue that, within 2 kpc, the inverse-parallax method gives comparable results (distances and sizes) as the Bayesian approach based on the exponentially decreasing volume density prior. Both of these methods show very similar dependence of the line-of-sight elongation of clusters (needle-like shapes resulting from the parallax uncertainties) on the distance. We also looked at a measure of elongations of the studied clusters and find the maximum distance of 665 pc at which a spherical fit still contains about half of the stellar population of a cluster. It follows from these results that the 3D structure of an open cluster cannot be properly studied beyond $\sim 500$ pc when using any of mentioned standard transformations of parallaxes to distances.
\end{abstract}

\section*{Introduction} \label{introduction}
Galactic star clusters are most important objects not only when it comes to describing the Milky
Way and its structure, but also for studying the individual stellar members. Various
star groups, such as variables and binaries, can be studied in star clusters in a statistical way. This is
based on the idea that the cluster reddening, age, distance and metallicity
can be assumed to be the same for each of the cluster members. These cluster parameters can be deduced by fitting proper isochrones, for example.

In the recent years, most open clusters were photometrically studied in a (semi-)automatic
way using 2MASS $JHK_{\mathrm{S}}$
and \textit{Gaia} $G_{\mathrm{BP}}$, $G_{\mathrm{RP}}$, and $G$ data. The traditionally photometric systems like the Johnson-Cousins
$UBVR_{\mathrm{C}}I_{\mathrm{C}}$ and Str{\"o}mgren $uvby\beta$ ones are hardly used any more. Especially critical is the lack
of observations in the ultraviolet region -- this makes it difficult to de-redden individual stars
or to get membership probabilities by using a classical $(U-B)$ versus $(B-V)$ diagram (Yontan et al. 2019), 
for example.

\begin{figure}[ht]
  \centering
  \includegraphics[scale=0.48]{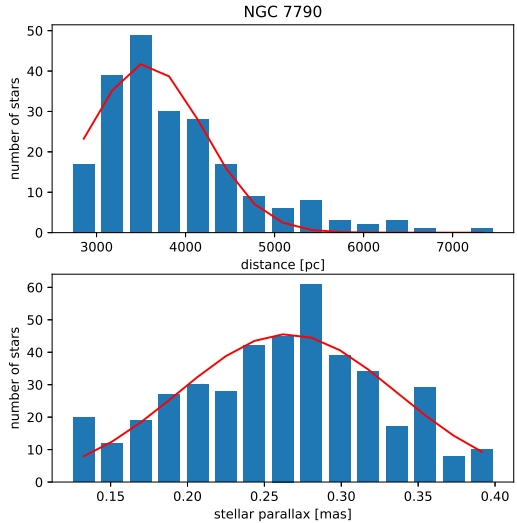}
  \caption{The distribution of the distances (upper panel) and parallaxes (lower panel) for all
members with probabilities higher than 50\% for NGC 7790 taken from Cantat-Gaudin et al. (2018).
The distances were calculated by inverting parallaxes. This approach is the reason why assembling a
symmetric parallax distribution always means to have a tail of much more distant members.}
  \label{NGC_7790} 
\end{figure}

With the launch of the \textit{Gaia} satellite the hopes were high to get precise membership
probabilities using parallaxes, proper motions, and radial velocities for a statistically
sound sample of star clusters. For the first time, even the intrinsic kinematics was hoped
to be investigated. The latter is important for the understanding how the angular momentum of
the initial molecular cloud is conserved during the formation and evolution of star 
clusters. We basically have no knowledge about the initial conditions when it comes to the rotational characteristics.
But for our understanding of kinematics and dynamics of a cluster (for example, modelling the first stages
of cluster evolution, studying the kinematic evolution of clusters and dynamical effects), this information is vital -- see, for example, 
K{\"u}pper et al. (2010) and Parker \& Wright (2016).

Cantat-Gaudin et al. (2018) presented a status report for 1229 open clusters on the basis of
the \textit{Gaia} DR2 release. They established a list of members and derived cluster parallaxes and distances within 
a given error range. The other three cluster parameters (age, reddening, and metallicity) 
were neither derived nor taken into account (especially the reddening). They also
reported the discovery of 60 new open clusters. These were identified
on the basis of consistent proper motions, parallaxes, and concentrations on the sky.
Using this method, Cantat-Gaudin et al. (2019) detected 41 additional new star clusters.
Later on, Monteiro et al. (2019) used also \textit{Gaia} photometry to get all four cluster
parameters for the above mentioned discovered aggregates using a cross-entropy global optimization 
algorithm to fit theoretical isochrones. However, their analysis showed that 80 candidates 
are likely not real open clusters. This already shows that kinematical data alone are not
sufficient and photometric data have to be taken into account when analysing star clusters.
Bossini et al. (2019) derived ages, reddening, and distances (for a fixed metallicity)
for 269 open cluster from the sample by Cantat-Gaudin et al. (2018). They have used an
automated Bayesian tool for fitting stellar isochrones to \textit{Gaia} photometry using the membership
probabilities from Cantat-Gaudin et al. (2018) for selecting the cluster sequences. Their sample
is biased because they selected only low reddening objects and discarded very young clusters.  
One of their main result is that 90\% of the clusters have a sigma of the absolute distance
modulus smaller than 0.037\,mag (median is 0.025\,mag). 
However, as they have shown, the errors increase by about one order of
magnitude when metallicities are taken into account. If all four cluster parameters (distance, extinction, age, and metallicity) are considered,
the differences of the derived values from different independent sources and data sets are quite large.
Fitting isochrones to an open cluster population is a complex procedure and depends, for example, on the turn-off
point and the location of the red giant population.
In the literature compilation by Netopil et al. (2015), the dispersion between different data sets amount to about 0.2 dex for the age,
0.08 mag for the reddening, and 0.35 mag for the distance modulus.
Similar or even larger discrepancies can be seen in Figs. 9 and 10 of Bossini et al. (2019). 

In this paper, we investigate limitations of the \textit{Gaia} DR2 data when it comes to the investigation
Galactic open clusters and their parameters. The paper is organized as follows: in Sect. \ref{selection}
we present the basic characteristics of the data set and the target cluster selection; in Sect. \ref{r_to_plx}
a summary of the problem of transforming parallaxes to distances is given; in Sect. \ref{analysis} 
we analyse in detail cluster distances and width.

\section*{1. Target cluster selection} \label{selection}
For our work, we used the data set based on the analysis by Cantat-Gaudin et al. (2018) who presented an unsupervised
membership assignment procedure to determine lists of cluster members based on the \textit{Gaia} DR2
catalogue. They provided the membership and mean parameters for a set of 1229 clusters and
401\,448 individual stars.
Their analysis is based on the membership assignment code
UPMASK (Unsupervised Photometric Membership Assignment in Stellar Clusters, Krone-Martins \& Moitinho 2014)
which does not rely on physical assumption about stellar clusters, apart from
the fact that its member stars must share common properties, and
be more tightly distributed on the sky than a random distribution. The analysis by Cantat-Gaudin et al. (2018) is based only on
kinematical and astrometrical data. 

One has to keep in mind that working in the parallax space is different than working in the distance
space (see Sect. \ref{distances_widths}). Basically, Cantat-Gaudin et al. (2018) used a maximum likelihood procedure, maximising
the probability of measuring the parallax of each individual star and 
the likelihood for the cluster distances to be the product of the
individual likelihoods of all its members. They neglected any correlations between parallax 
measurements of all stars such as suggested in Sch{\"o}nrich et al. (2019), for example. The resulting 
distribution should be symmetrical for the parallaxes, but is not
for the distances. In Fig.~\ref{NGC_7790}, the example of this effect for NGC 7790 is shown. 
This is not unexpected - Luri et al. (2018) already showed this
effect in their analysis of the \textit{Gaia} DR2 data. Working with a symmetric parallax distribution always means to
have a tail of much more distant members for the investigated cluster (when using inverse parallax approach). 
In the case of NGC 7790 (distance of about
3700 pc) this means that members are distributed from 2700 to 7500 pc with about 15\% of stars more distant than
4700 pc. These distances were calculated using $r=\varpi^{-1}$ for the individual cluster members. The conversion from
parallaxes to distances is clearly problematic. However, it will be shown in Sect. \ref{r_to_plx}
that there is a simple statistical solution when dealing with open clusters.

In the following, we define two different samples for which we did our analysis. These samples are
defined as 

\begin{itemize}
\item ``Loose sample'': number of stars in a cluster $>$\,50, individual parallax (or distance) error $<$\,50\%, and individual membership probability $>$\,50\%; 938 aggregates
\item ``Strict sample'': number of stars in a cluster $>$\,300, individual parallax (or distance) error $<$\,5\%, and individual membership probability $>$\,70\%; 181 aggregates
\end{itemize}

The individual parallax/distance error refers to the fact that we will begin our analysis with both, starting in the parallax space
and in the distance space. The mentioned errors are taken from the used data sets.

For the purpose of this work, we have chosen to work with the data from Cantat-Gaudin et al. (2018) and Bailer-Jones et al. (2018).
The former presents one of the most recent compilations of parallaxes for a larger number of open clusters. The latter
data set gives the largest sample ($\sim 10^9$) of distances for Galactic stars -- we have used those which coincide with the catalogue from Cantat-Gaudin et al. (2018).
Although these distances may not be the best for analysing open clusters, they present a good starting point and an option for a comparison with a different
approach.

The discrepancy of the number of clusters for the loose sample and the total number is explained by the fact
that almost 300 open clusters have less than 50 members when applying the individual distance errors and membership probabilities.
We have chosen the upper limit of 50 members because otherwise the distance distributions in the histograms are
mostly dominated by noise which creates problems for the fitting procedure described in the next sections.

We also have to emphasize that 1795 stars were found to be members of at least two open clusters
(49 individual ones in total), 579 of them with a membership probability of higher than 50\% for both clusters.
Although this number is insignificant compared to the overall number of investigated stars, it still shows
that there are shortcomings in the numerical procedure for deriving the cluster memberships.

\section*{2. Calculation of distances from parallaxes} \label{r_to_plx}

As was mentioned, calculating the distance $r$ by inverting the parallax $\varpi$ is a problematic approach (for more details, see Luri et al. 2018).
This is due to the fact that the measurement is accompanied by an uncertainty. If we assume that the probability density function (PDF) for a parallax measurement is a normal distribution it will not transform to a normal distribution by assuming $\varpi^{-1}$. Instead, it will produce a longer tail towards the larger distances and the maximum of the distribution will be located at somewhat shorter distance when compared with the true distance. This effect will increase with the value of the relative error $f=\frac{\sigma_\varpi}{\varpi}$ of the parallax measurement. Obviously, for some (small) values of $f$ the differences become quite negligible.

In this section, we aim to examine how this affects the determination of distances toward open star clusters. Although inferring precise values of distances from parallaxes can be quite complicated for the field stars (due to the transformation problems discussed above), we believe that the situation gets much better for open clusters (although not for the individual cluster members).

\subsection*{2.1. Getting distances toward open clusters} \label{r_to_plx_1}

\begin{figure}[t]
  \centering
  \includegraphics[scale=0.5]{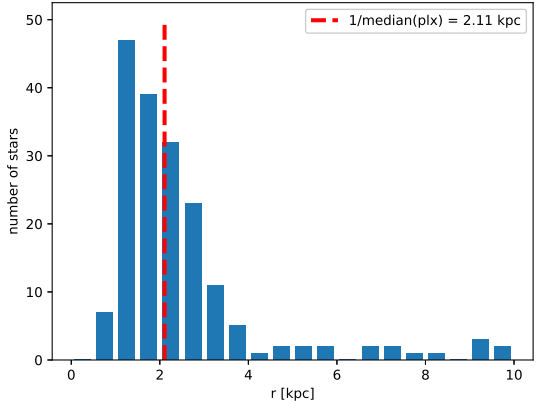}
  \caption{Simulation of an open cluster located 2.00 kpc away. The red dashed line indicates the determined distance of a cluster calculated as the inverse of the median of all parallaxes of the cluster members.}
  \label{fignew1}
\end{figure}

Due to the fact that the observed number of members of a cluster represents a number of distinct measurements of the centre of the cluster (in terms of parallaxes), the distribution of parallaxes should represent the PDF in parallax space. Finding the maximum of this distribution (usually close to a normal distribution) is very simple and can be achieved by calculating the median or fitting a Gaussian function. Since the centre of the distribution is assumed to be the true parallax $\varpi_{\mathrm{true}}$, the inverse of this value should yield the true distance $r_{\mathrm{true}}$ towards the cluster.

Similar to Luri et al. (2018), we have simulated a cluster of 200 stars at the distance $r_{\mathrm{true}}=2.0$ kpc using a normal distribution with scale $s=5$ pc (representing the cluster size). Then we transformed these 200 true distances to true parallaxes using $\varpi_{\mathrm{true}} = r_{\mathrm{true}}^{-1}$. The observed parallaxes were generated using normal distribution with the centre located at the true parallaxes and assuming the width of the distribution to be $\sigma_{\varpi}=0.3$ mas. From those, we could construct the distribution of distances of the individual cluster members using a Bayesian approach. However, if we are only interested in calculating the distances toward the centres of the studied clusters, this may be unnecessary.

Instead, we are going to look at the distribution of parallaxes. Assuming that the distribution is normal, we can fit the data with a Gaussian which will give us the parameters of the fit, $\varpi_0$ and $\sigma_{\varpi}$, together with their uncertainties. If the number of the observed stars of a cluster is high enough and the observational uncertainty is low enough for the most of the members, we can safely assume that $\varpi_{\mathrm{true}}=\varpi_0$ and determine the distance towards the centre of the cluster (Fig.~\ref{fignew1}).

\begin{figure}[t]
  \centering
  \includegraphics[scale=0.5]{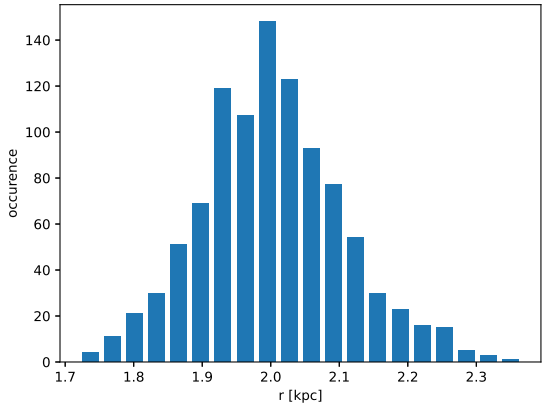}
  \caption{Based on 1000 simulations of the same cluster, we can generate a histogram of the distances found by the inverse of the $\textrm{median}(\varpi)$ (as in Fig.~\ref{fignew1}). The centre of this distribution is located at $R=2.00$ kpc with $\sigma_R=0.11$ kpc.}
  \label{fignew2}
\end{figure}

To test the results of this method, we can generate the same cluster (with the same input parameters mentioned above) a thousand times and can construct a histogram of the found distances. Analysis of this histogram (Fig.~\ref{fignew2}) yields us the variance of the distances found using this method. The variance depends on the input parameters -- the number of cluster members and the observational uncertainties. For our test case, we find the values of median $R=2.00$ kpc and the standard deviation $\sigma_R=0.11$ kpc. It is worth pointing out that this gives about an order of magnitude smaller relative error than we have for the assumed measurement uncertainty.

\subsection*{2.2. Determining the distance errors} \label{r_to_plx_2}

In reality, we do not have a thousand of observations of all members for the same cluster. However, we can still learn much from the Gaussian we used to fit the observed parallaxes of a cluster. All of the information about the variance of the final distance is found in the parameter which describes the centre of the Gaussian. It is therefore a good first approximation to use the uncertainty of this parameter to determine the uncertainty of the final value of the distance. This can be calculated from the simple approach as $e_r = \frac{e_{\varpi_0}}{\varpi_0^2}$, assuming that $\frac{e_r}{r} \ll f$. If we assume that $r \approx R$ and $e_r \approx \sigma_R$ ($r$ represents the distance we get from one simulation of a cluster, $R$ is the distance we get from one thousand simulations of the same cluster), then the term on the left hand side tends to be about an order of magnitude smaller than $f$, so this should be a good assumption. We present the values of distances $r$ and $R$ for 6 simulated clusters in Table \ref{sims_distances}. We have used different numbers of members and different observational uncertainties.

As we can see, the results are, for the most part, quite similar. The exceptions are such clusters where the number of the observed cluster members is lower than $N \sim 100$ and the observational relative error is higher than $f \sim 0.50$. Moreover, the distance towards the cluster plays a crucial role. It is also worth mentioning that the calculated uncertainties of the fit parameters of a single cluster will slightly vary due to the randomness included in the cluster generation procedure.

\begin{table*}[t]
\caption{Comparison of the true distances with the distances $r$ and $R$ found by inverting $\varpi_0$ for six different simulated clusters of size $s=5$ pc. As expected, $R$ very closely matches the true distances of a given cluster. Although the values of $r$ typically differ from $R$, the difference $|r-R|$ is usually lower than $e_r$.}
\label{sims_distances}
\begin{center}
\begin{tabular}{c|ccccccc}
\hline     
Cluster & $N$ & $r_{\mathrm{true}}$ & $f$  & $r$ & $e_r$ & $R$ & $\sigma_R$ \\
 &  & {[}kpc{]} & & {[}kpc{]} & {[}kpc{]} & {[}kpc{]} & {[}kpc{]} \\
\hline
1       & 200 & 2.0           & 0.60 & 2.12            & 0.13                 & 2.00                 & 0.11                      \\
2       & 50  & 2.0           & 0.60 & 2.22            & 0.41                 & 2.00                 & 0.22                      \\
3       & 50  & 2.0           & 0.20 & 2.05            & 0.05                 & 2.00                 & 0.07                      \\
4       & 200 & 4.0           & 0.60 & 3.82            & 0.12                 & 3.99                 & 0.21                      \\
5       & 50  & 4.0           & 0.40 & 4.26            & 0.46                 & 3.99                 & 0.27                      \\
6       & 100 & 5.0           & 0.25 & 5.21            & 0.15                 & 5.01                 & 0.16                    \\
\hline           
\end{tabular}
\end{center}     
\end{table*}

Generally, if the number of cluster members is $N>50$ and the observational relative error is $f<0.50$ then we can use the described procedure to determine distances (and their uncertainties) toward open clusters quite precisely up to $r_{\mathrm{true}} \sim 4$ kpc. However, it should be possible to use this approach also for the more distant clusters with $N > 100$ if the observational error is lower than $f \sim 0.25$.

\subsection*{2.2. Variations in the observational uncertainties} \label{r_to_plx_3}

Unfortunately, the parallax measurement uncertainties are not the same for all members of a cluster. Let us take a look at what happens when we assume a distribution of uncertainty values.

\begin{figure}[t]
  \centering
  \includegraphics[scale=0.5]{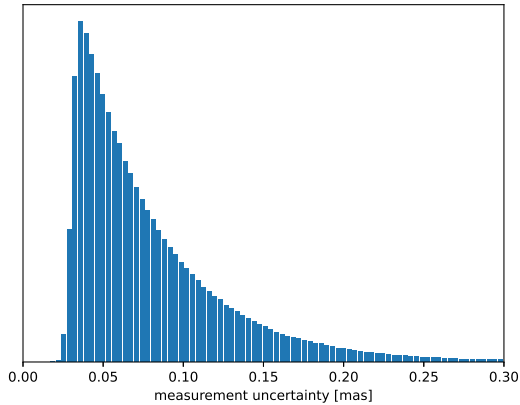}
  \caption{Estimated probability density function for the uncertainty of the parallax measurement. The function is composed of two parts -- the normal distribution creates a short tail toward the lower values and defines the position of the peak of the PDF (does not coincide with the peak of the normal distribution), the exponential distribution creates the long tail toward the higher values of uncertainty. The probability that the uncertainty will be lower than 0.32 mas is about 99.7 \%.}
  \label{fignew3}
\end{figure}

We can find in the data from Cantat-Gaudin et al. (2018) that for distances below 3 kpc the distribution of uncertainties can be well described by a combination of a Gaussian distribution together with an exponential distribution. The probability distribution we used can be seen in Fig.~\ref{fignew3}. The Gaussian in the distribution is required to produce the short tail towards the smaller values. Finally, it must be mentioned that for clusters beyond 3 kpc the position and width of the Gaussian term increase with the distance. We have decided to ignore this small discrepancy at larger distances since we are only interested in the effect that such a distribution has on the determined distance errors.

Let us simulate two very different situations ($N=50$ and $N=200$) to study the effect of the distribution of uncertainties. We have generated 80 clusters between 0.05 kpc and 4.00 kpc (with equidistant steps) and for each cluster we calculate $\frac{\sigma_R}{R}$. We want to know how this ratio varies with the number of cluster members and with distances. If the distribution of $\sigma_{\varpi}$ plays any important role, it should present itself as a spread in the curve of the plot of the relative error of $R$ against the true distances.

\begin{figure}[t]
  \centering
  \includegraphics[scale=0.5]{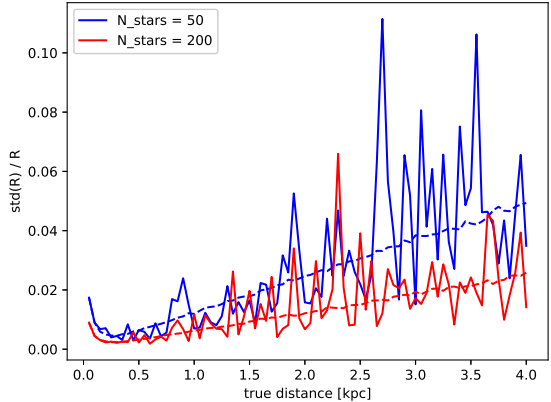}
  \caption{The influence of the distribution of measurement uncertainties on $\sigma_R / R$. Full lines represent the values calculated with the assumed distribution of uncertainties, dashed lines represent the same cluster but assuming constant $\sigma_{\varpi}=0.07$ mas.}
  \label{fignew4}
\end{figure}

As we see in Fig.~\ref{fignew4}, the ratio $\frac{\sigma_R}{R}$ gets (statistically) smaller for larger values of $N$. Also, it increases with the true distance of the cluster, as was expected. However, it is apparent that the plot is dominated by the variation of the observational uncertainty which produces the scatter around the dashed lines (the same as full lines, but with fixed uncertainty $0.07$ mas). It should be noted that only rarely $\frac{\sigma_R}{R}>0.10$ with the assumed distribution of $\sigma_{\varpi}$.

\begin{figure}[t]
  \centering
  \includegraphics[scale=0.5]{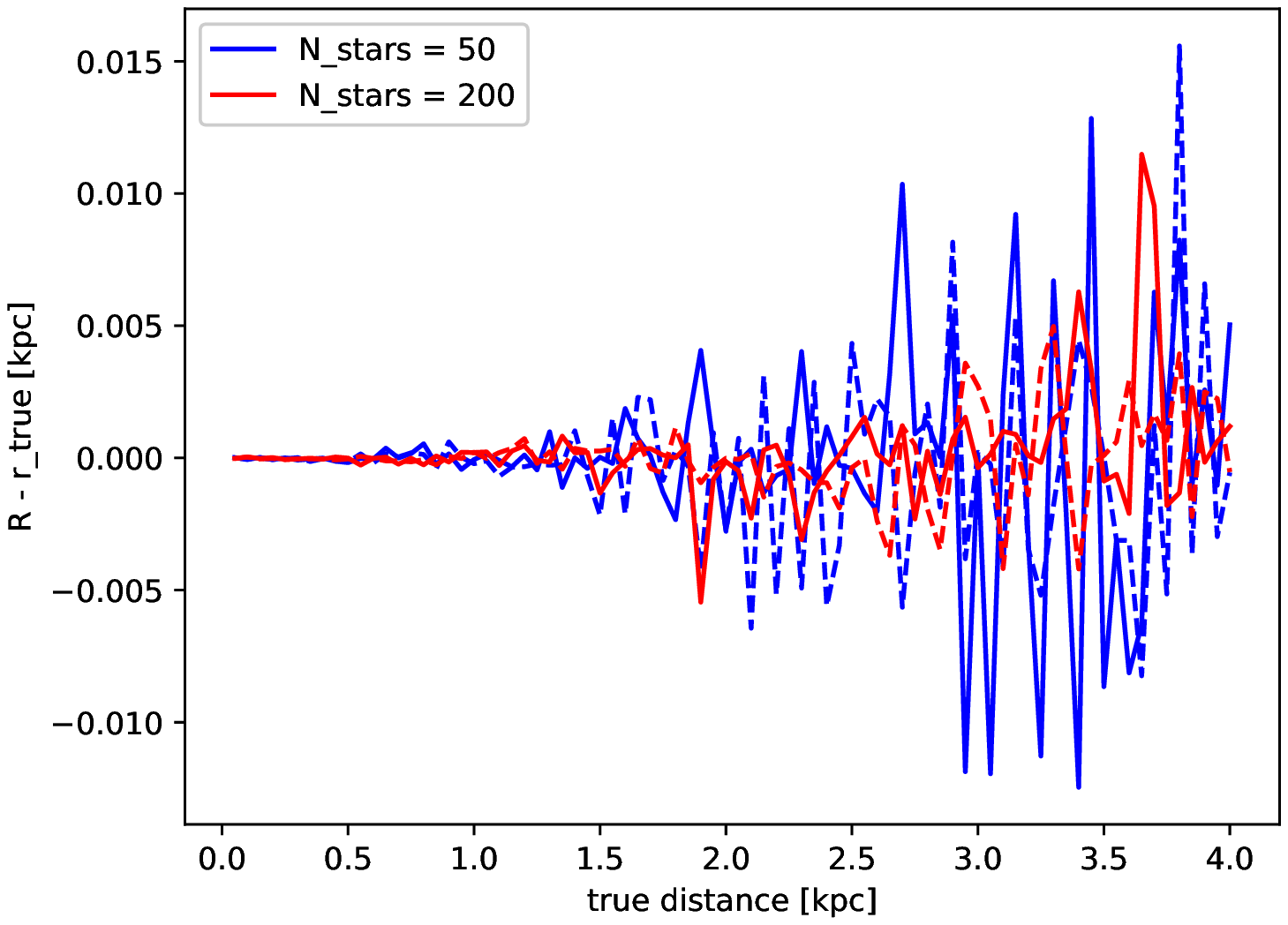}
  \caption{The influence of the distribution of measurement uncertainties on $R - r_{\mathrm{true}}$. Full lines represent the values calculated with the assumed distribution of uncertainties, dashed lines represent the same cluster but assuming constant $\sigma_{\varpi}=0.07$ mas.}
  \label{fignew5}
\end{figure}

We can also display the deviations of the values of $R$ from the true distances $r_\mathrm{true}$ using the same approach but putting $R-r_\mathrm{true}$ on the $y$-axis. This is shown in Fig.~\ref{fignew5}. Statistically speaking, these differences are very small, therefore we assume that the designed procedure should work very well assuming that the typical observational uncertainty is low enough and that the number of cluster members is sufficient (this is a very good assumption for our ``strict sample'').

\subsection*{2.4. Individual cluster members} \label{r_to_plx_4}

The last question that remains to be answered -- can we calculate the distances toward the individual stars of a cluster with the procedure described above? Unfortunately, the answer is no. To calculate the individual distances one must rely on the Bayesian approach assuming a reasonable prior probability distribution and then determine the posterior distribution. Possible approaches to this problem were described by Bailer-Jones (2015). Another solution would be to observe the parallax of the same star multiple times but this is not practical.

Since the goal of this work is not the calculation of the distances but rather the statistical analysis of the effects of observational uncertainties on the results derived from \textit{Gaia} DR2, we will look at this problem from a different perspective. As we already have an acceptable estimate of the distance and of its error ($r$, $e_r$), we only need to get the variances of distances of the individual members of clusters. To find these, we will use the distances $r_{\mathrm{bay}}$ resulting from a Bayesian approach (taken from Bailer-Jones et al. 2018) and the inverse-parallax distances $r_{\mathrm{inv}}$.

Before we start our analysis, let us estimate what we should expect from $r_{\mathrm{bay}}$ and $r_{\mathrm{inv}}$. We will simulate a cluster of 1000 stars at the distance $r_{\mathrm{true}}=2.0$ kpc with cluster size $s=5$ pc. Inverse-parallax distances toward the individual stars are calculated simply as $r_{\mathrm{inv}}=\varpi^{-1}$. The Bayesian distances can be calculated as described in Bailer-Jones (2015), with the help of the exponentially decreasing volume density prior -- the distance of a star is found as the maximum in the PDF assuming the characteristic length scale $L=1.2$ kpc.

\begin{figure}[t]
  \centering
  \includegraphics[scale=0.5]{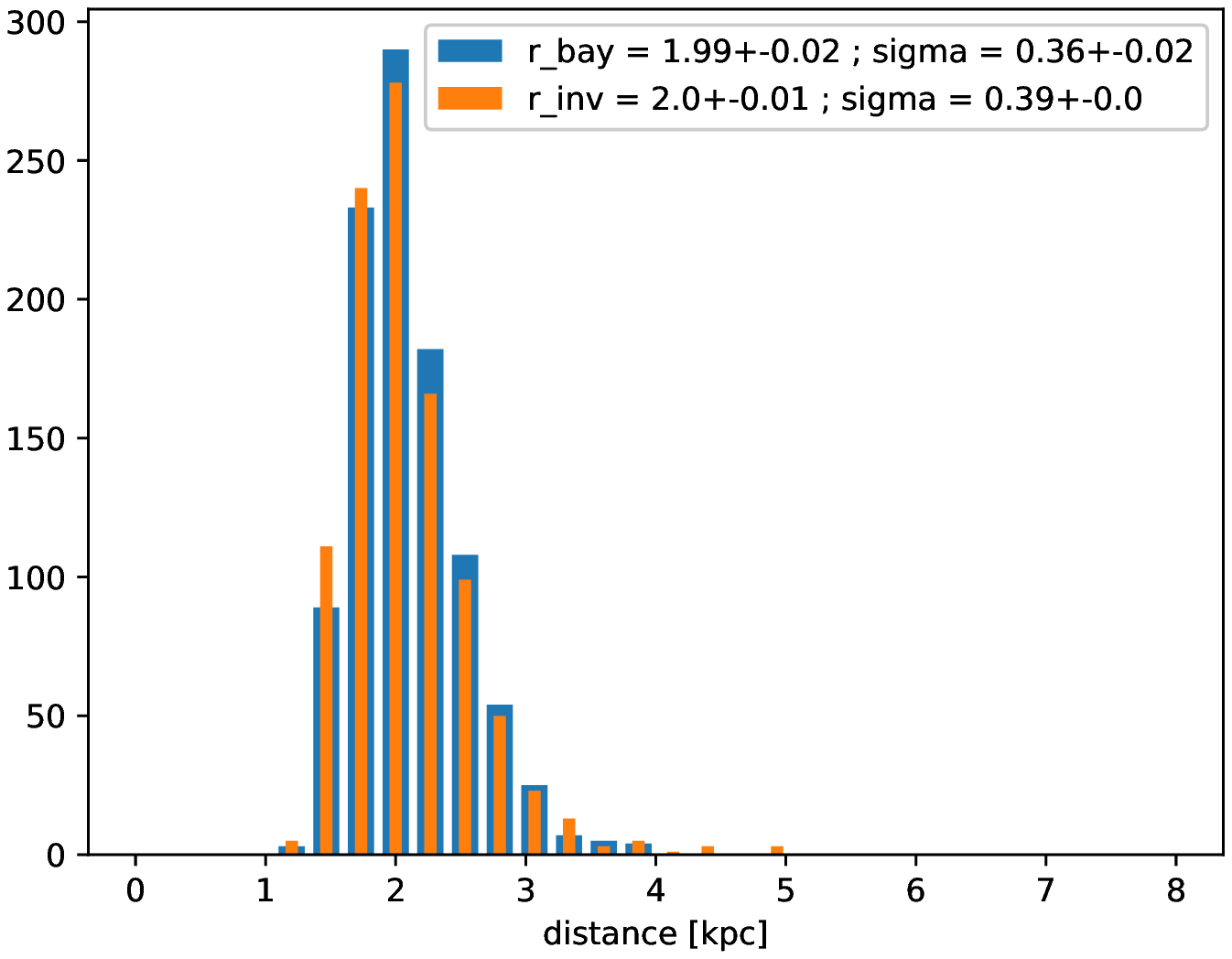}
  \caption{Comparison of the Gaussian parameters derived from Bayesian distances $r_{\mathrm{bay}}$ and inverse-parallax distances $r_{\mathrm{inv}}$, assuming $\sigma_\varpi = 0.10$ mas. In the approach of inverse-parallaxes, the distribution of parallaxes was used instead of the distribution of $r_{\mathrm{inv}}$. The used method is described in Sects. \ref{r_to_plx_1} and \ref{r_to_plx_2}.}
  \label{fignew6}
\end{figure}

\begin{figure}[t]
  \centering
  \includegraphics[scale=0.5]{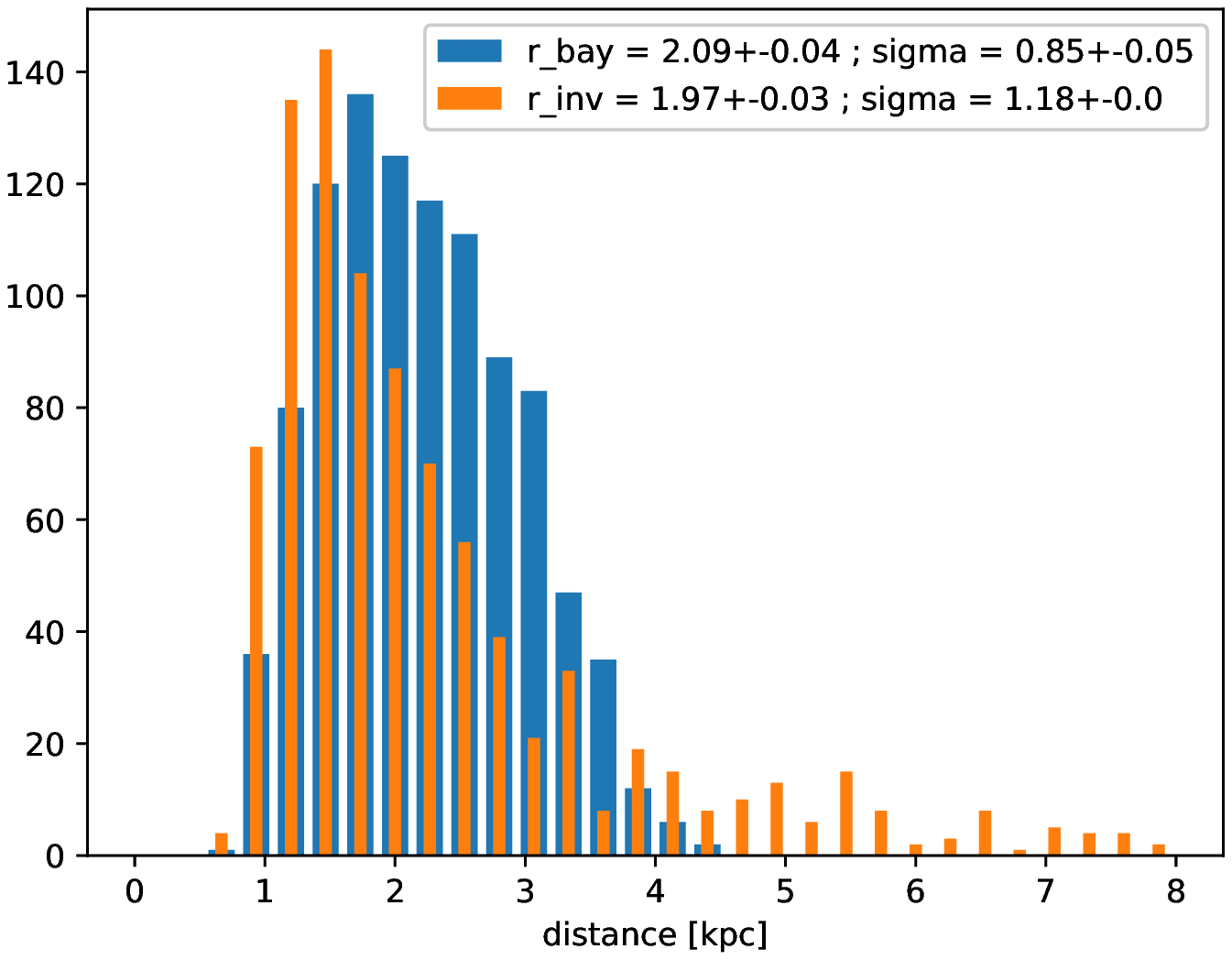}
  \caption{Same as Fig.~\ref{fignew6} but assuming $\sigma_\varpi = 0.30$ mas.}
  \label{fignew7}
\end{figure}

To characterize the cluster, we would like to fit the distributions of both distance measures with Gaussians (for simplicity) and use their parameters for describing the results. However, in the case of $r_{\mathrm{inv}}$ the distributions may be quite asymmetrical at large cluster distances (or with large relative observational uncertainties). Therefore, we have decided to fit $r_{\mathrm{bay}}$ with a Gaussian directly, while in the case of $r_{\mathrm{inv}}$ we first stay in the parallax space. Parallax distribution is fitted by a Gaussian (as was discussed in Sects. \ref{r_to_plx_1} and \ref{r_to_plx_2}), and we use the width of this distribution as the measure of the width of the distribution in distances as $\sigma_r = \frac{\sigma_{\varpi}}{\varpi_0^2}$. This measure should not be confused with the mentioned calculation of the distance errors in Sect. \ref{r_to_plx_2} -- in this case, the relative errors are clearly very similar ($\frac{\sigma_\varpi}{\varpi_0} = \frac{\sigma_r}{r}$), therefore we need to be very careful when assigning a meaning to the quantity $\sigma_r$.

For $\sigma_\varpi=0.10$ mas, the widths of these distributions are almost identical ($\frac{\sigma_{\mathrm{inv}}}{\sigma_{\mathrm{bay}}} \sim 1.10$, see Fig.~\ref{fignew6}). On the other hand, when we take $\sigma_\varpi=0.30$ mas, the Bayesian approach starts giving significantly smaller Gaussian widths ($\frac{\sigma_{\mathrm{inv}}}{\sigma_{\mathrm{bay}}} \sim 1.45$, see Fig.~\ref{fignew7}). However, it should be noted that such high observational uncertainties are rare in our two data samples and the typical relative uncertainty never gets close to 50~\%.

Clearly, Bayesian approach is superior to the naive inverse-parallax approach. However, both methods give very similar values of distribution widths (in terms of the widths of fitted Gaussians), significantly deviating only in the most extreme cases. Based on this, we expect the two approaches to give approximately the same values of cluster widths.

\section*{3. Analysis} \label{analysis}
In the following subsections we present a detailed astrometrical and kinematical analysis of the targets.
Notice that throughout the paper the errors in the final digits of the corresponding quantity are given in parentheses.

\subsection*{3.1. Distances and widths} \label{distances_widths}

To start with, we will define the rectangular Galactic coordinates [$X\,Y\,Z$] using the spherical Galactic coordinates $r$, $l$, and $b$

\begin{figure}[t]
\centering
\includegraphics[scale=0.50]{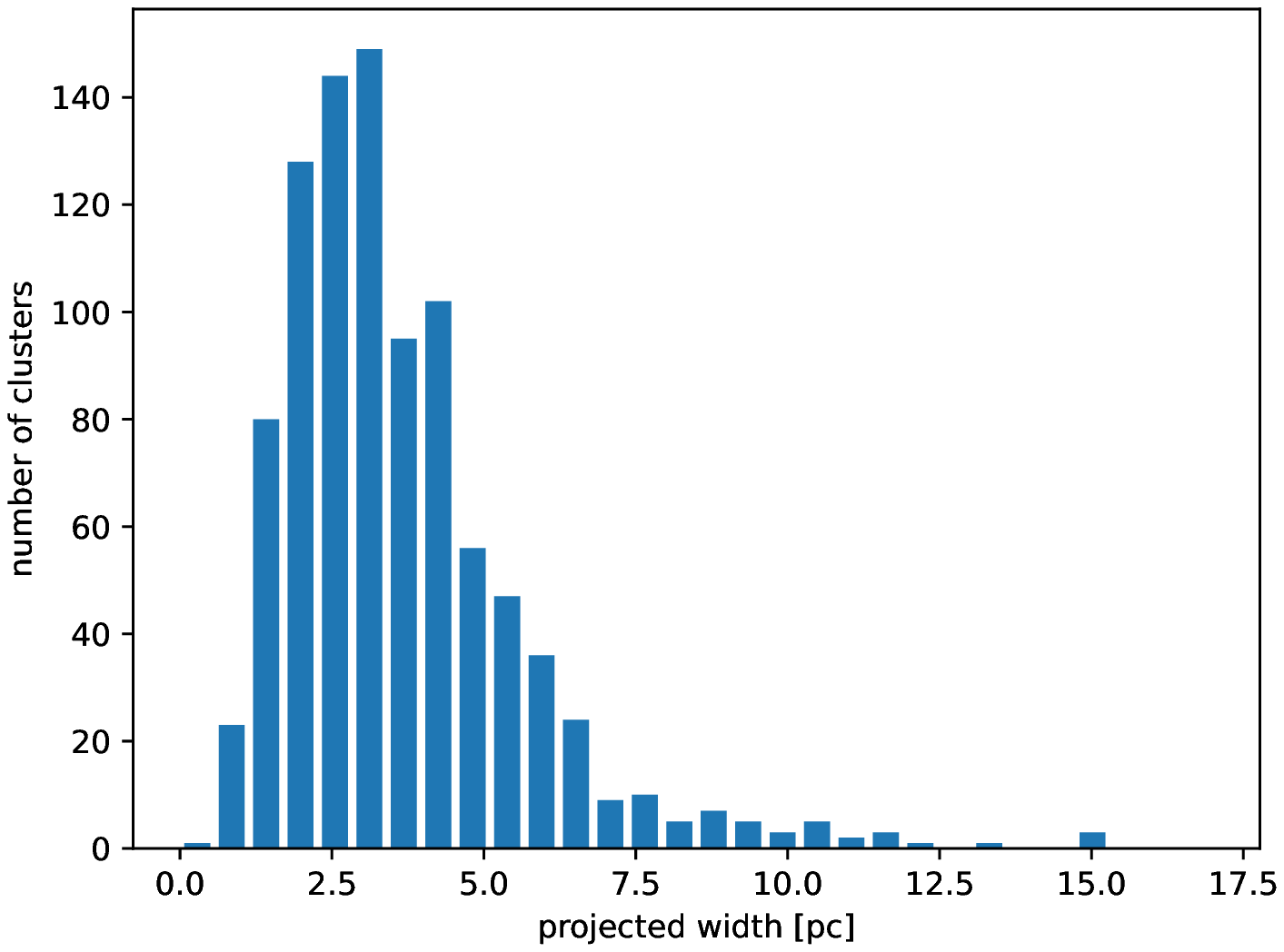}
\caption{The histogram of the projected widths. We can interpret 
these widths as a measure of the cluster size or diameter.}
\label{widths_OCLs} 
\end{figure}

\begin{eqnarray}
X &=& r \, \cos{l} \, \cos{b} \,\,,\\
Y &=& r \, \sin{l} \, \cos{b} \,\,,\\
Z &=& r \, \sin{b} \,\,.
\end{eqnarray}
Assuming spherical symmetry on the sky, we need to pick only two coordinates to characterize a cluster in a spatial 3D space. The first of these coordinates is the projected distance $d_1$ which can be calculated by projecting the distance $r$ of the star on the vector oriented in the line-of-sight towards the centre of the cluster

\begin{equation}
d_1 = r \, \cos(\phi)
\end{equation}
where $\phi$ is the angle between the reference vector (centre of the cluster) and the vector oriented towards 
the given star. It should be noted that the value of $d_1$ is almost indistinguishable from the value of $r$, therefore we will use $r$ as the measure of the distance throughout this work. The second coordinate is the projected width $d_2$, which can be calculated similarly

\begin{equation}
d_2 = r \, \sin(\phi).
\end{equation}

As was mentioned in Sect. \ref{r_to_plx}, we cannot easily analyse a cluster with the distances toward individual cluster members. Instead, if possible, we would like to use the whole collection of stars to determine some properties of a given cluster. To do this, first we have to calculate the distance towards the cluster $r$. Then we can analyse the distribution of the Galactic coordinates $l$ and $b$ which are determined quite precisely and determine the angular size of a cluster $\phi_{\mathrm{clust}}$ as the mean of the standard deviations of the two coordinates. The absolute cluster size can then be determined as the projected width $d_2$. If this is done for each of the clusters in our samples, we can find the typical spatial diameter of a cluster by analysing the distribution of the projected widths (which can be done by calculating the median and the median standard deviation, for example).

We have applied the procedure discussed in Sects. \ref{r_to_plx_1} and \ref{r_to_plx_2} to our samples and calculated
their distances and projected widths, together with the corresponding errors. For the most part, we will
focus our attention on the loose sample. The only aggregate from the complete sample that we were not able to fit in the distance space is Tombaugh 2
which is a distant (6 to 8\,kpc from the Sun) old 
open clusters in the direction of the Galactic anti-centre Cantat-Gaudin (2016). The available studies agree that its metallicity 
is sub-solar with a widespread of given values ($-$0.07 to $-$0.44\,dex). Looking into WEBDA\footnote{https://webda.physics.muni.cz}  
there is no doubt that this is a true open cluster, but the colour-magnitude diagram presented in Fig.~2 by Cantat-Gaudin et al. (2018) 
shows several different main sequences probable due to foreground stars.

\begin{figure}
\centering
\includegraphics[scale=0.35]{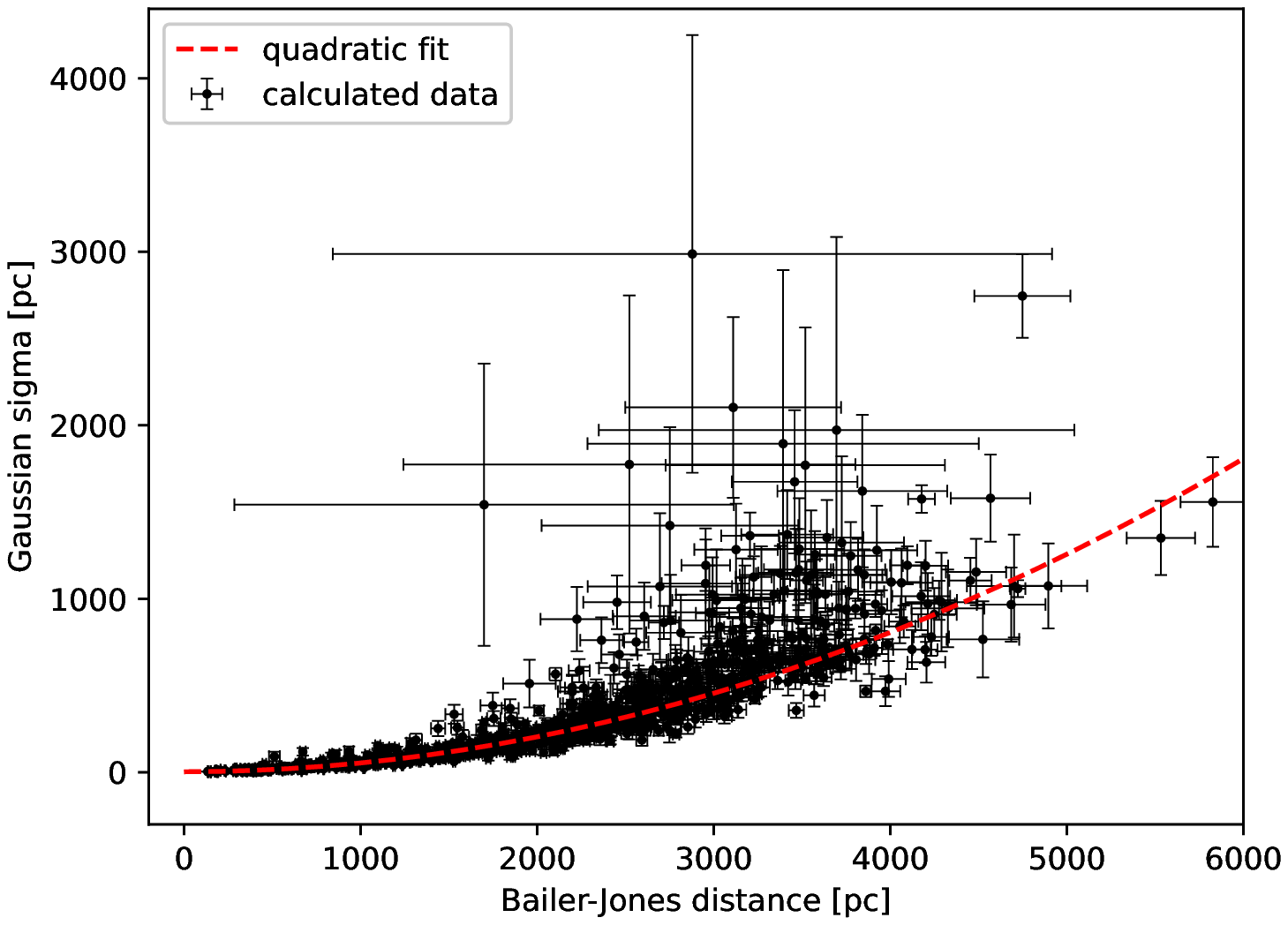}
\includegraphics[scale=0.35]{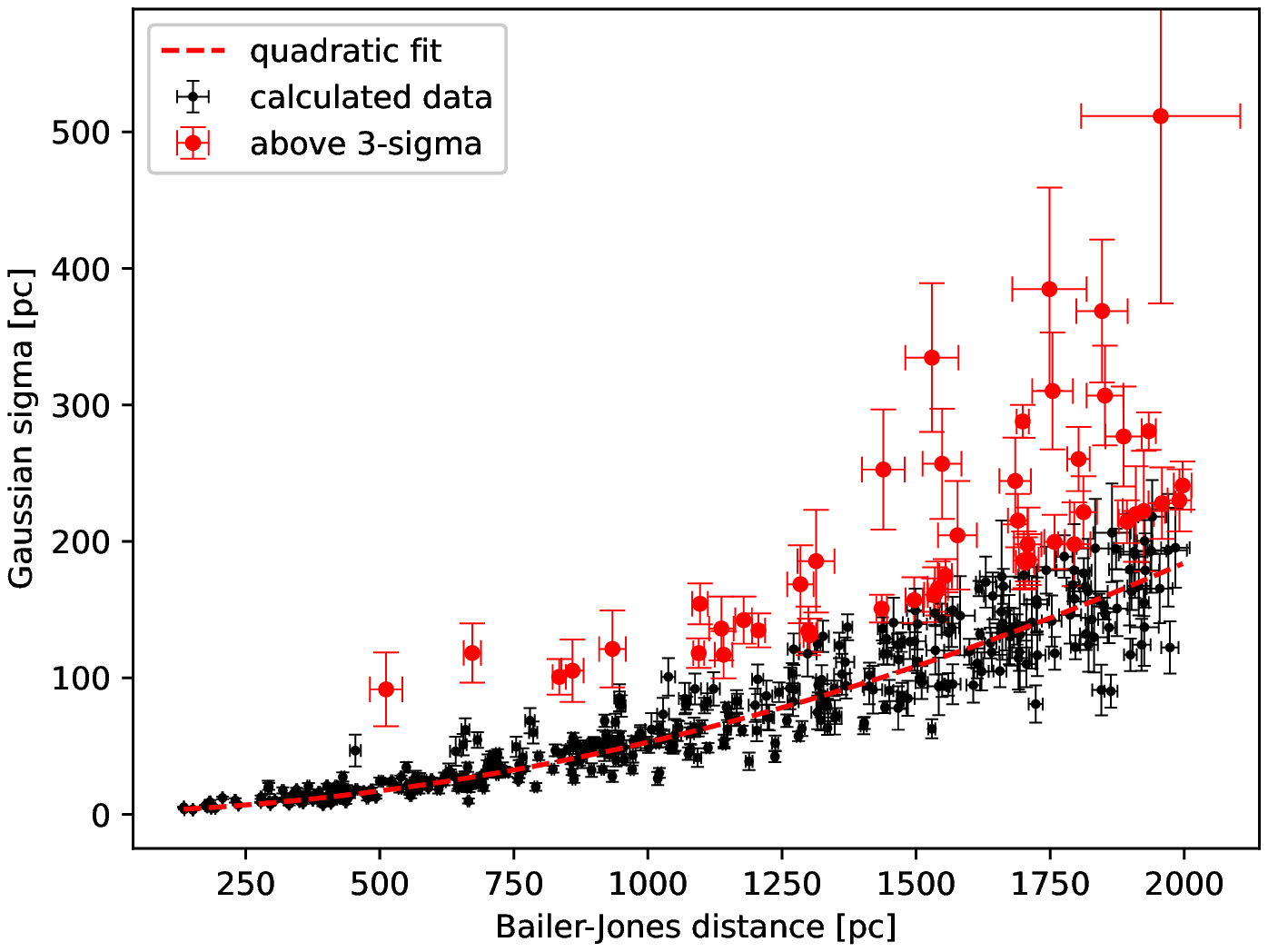}
\caption{The dependency of the sigma parameter on the Bailer-Jones distances for the loose
sample (upper panel) and limited to 2\,kpc (lower panel), respectively. The red circles
in the lower panel are the 50 clusters deviating more than 3$\sigma$ from the quadratic relation
(red line).}
\label{distance_sigma1} 
\end{figure}

\begin{figure}
\centering
\includegraphics[scale=0.35]{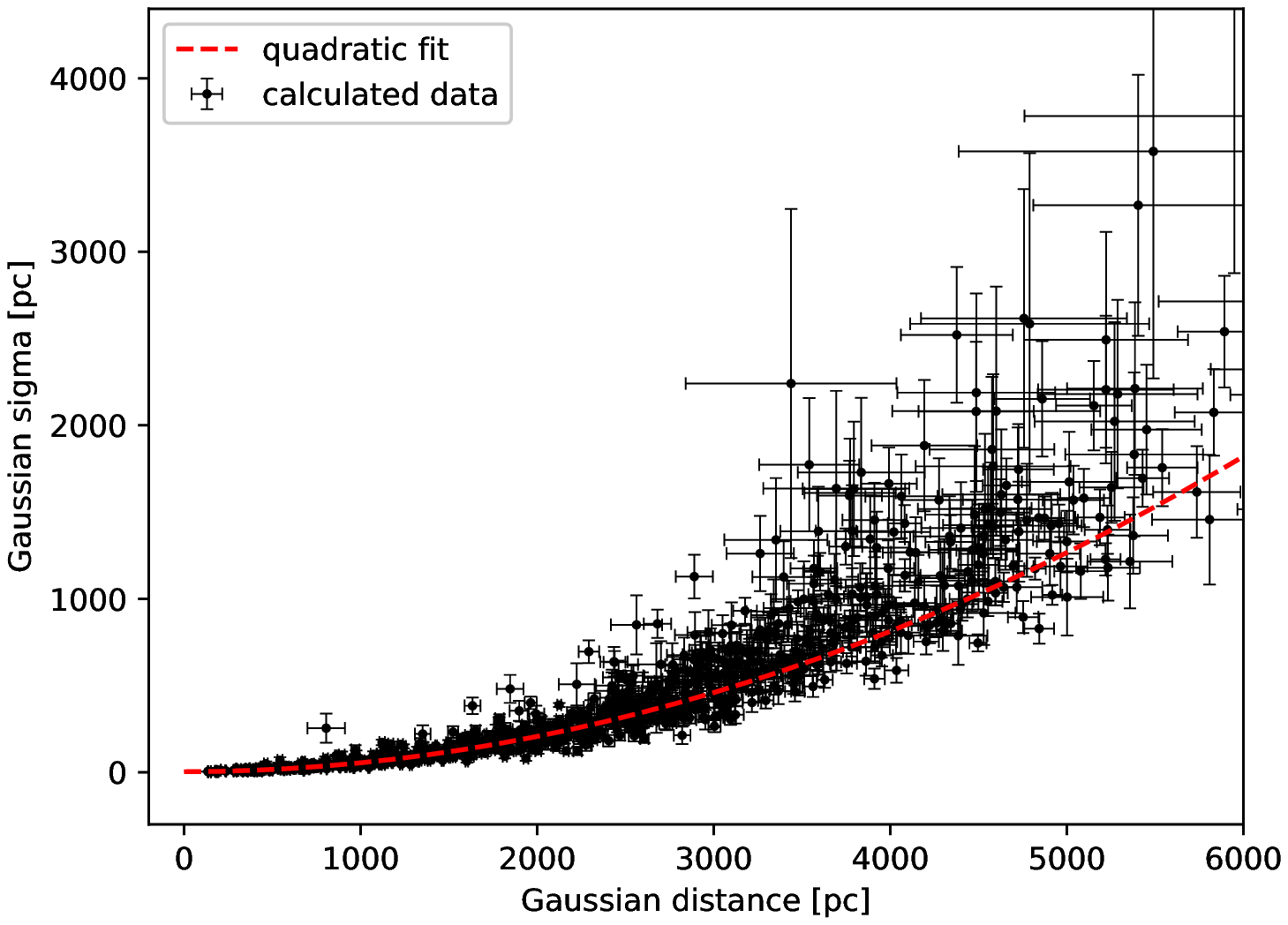}
\includegraphics[scale=0.35]{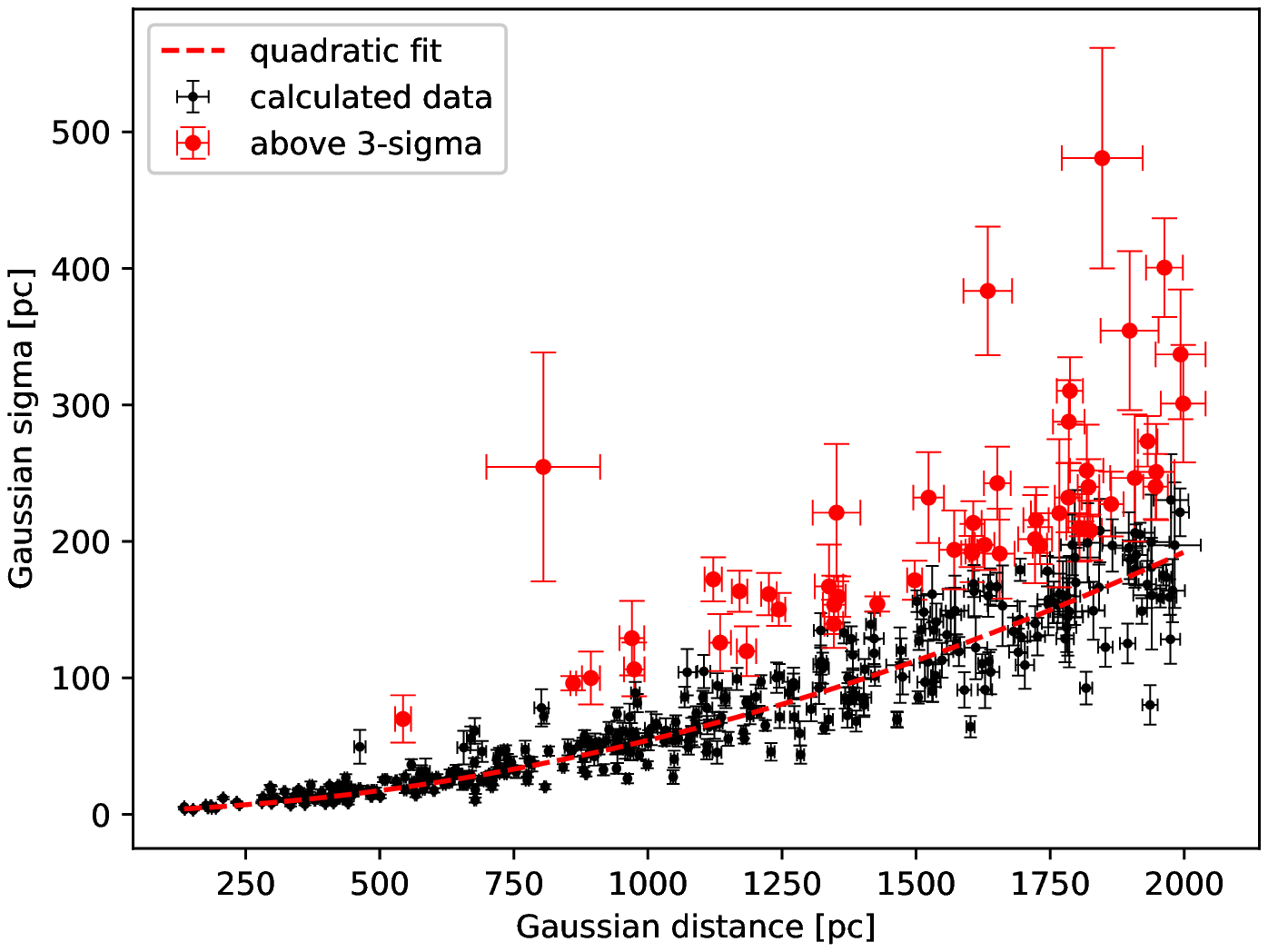}
\caption{The dependency of the sigma parameter on the inverse-parallax distance for the loose
sample (upper panel) and limited to 2\,kpc (lower panel), respectively. The red circles
in the lower panel are the 50 clusters deviating more than 3$\sigma$ from the quadratic relation
(red line).}
\label{distance_sigma2} 
\end{figure}

As the next step, we can analyse the distribution of projected widths. It should be 
noted that many clusters in this data set display different behaviour than normal -- however, a Gaussian 
fit is still a viable option.  It is also interesting to note in the histogram of those values (Fig.~\ref{widths_OCLs}) -- we 
see that the projected width of the clusters peaks at about 2.7(1.6)\,pc. When using the strict sample, the peak is located at 3.0(1.3)\,pc.
This result is comparable to the 2D analysis of the cluster sizes Kharchenko et al. (2013) which 
tells us that the typical size of a cluster is of the order of several parsecs.  
Moreover, the upper limit of absolute cluster
diameters of about 25\,pc (van den Bergh 2006) is also nicely supported. This upper limit is caused by the dimension 
of the initial molecular cloud from which the clusters are formed and the dissipation due to the differential rotation
of the Milky Way (Joshi et al. 2016). 

\begin{figure*}
   \centering
	 \begin{tabular}{cc}
   \subfloat[][NGC 1039 at $d$\,=\,505(2)\,pc. \label{NGC_1039}]{\includegraphics[scale=0.48]{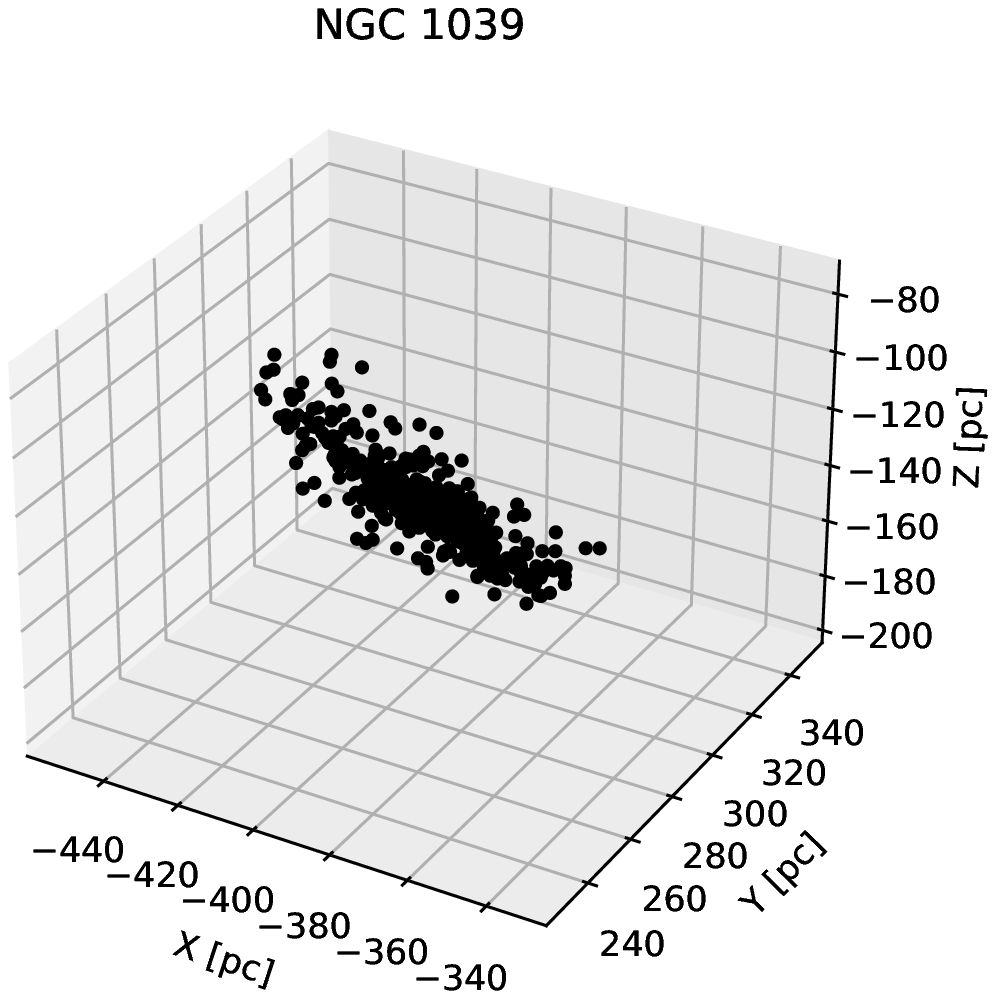}} \quad &
	 \subfloat[][NGC 1528 at $d$\,=\,1021(4)\,pc. \label{NGC_1528}]{\includegraphics[scale=0.48]{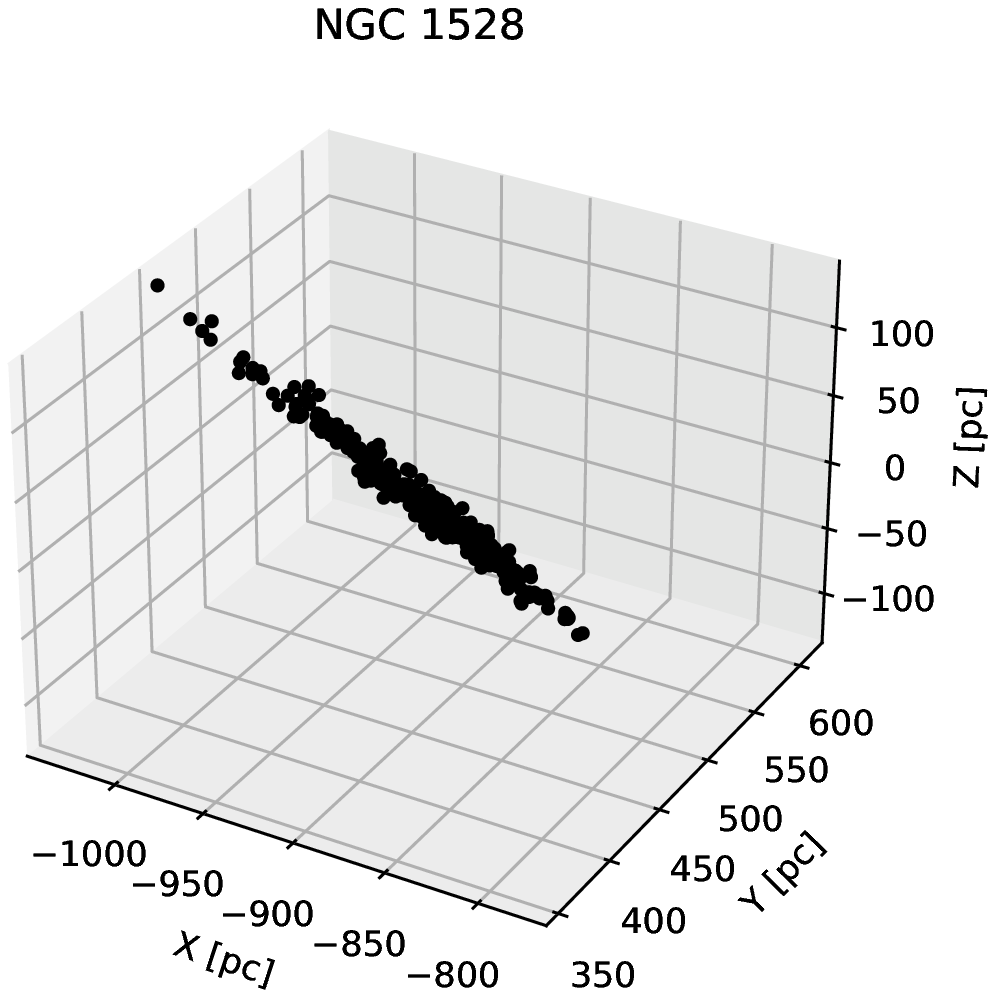}} \\
   \subfloat[][NGC 2632 at $d$\,=\,186($<1$)\,pc. \label{NGC_2632}]{\includegraphics[scale=0.48]{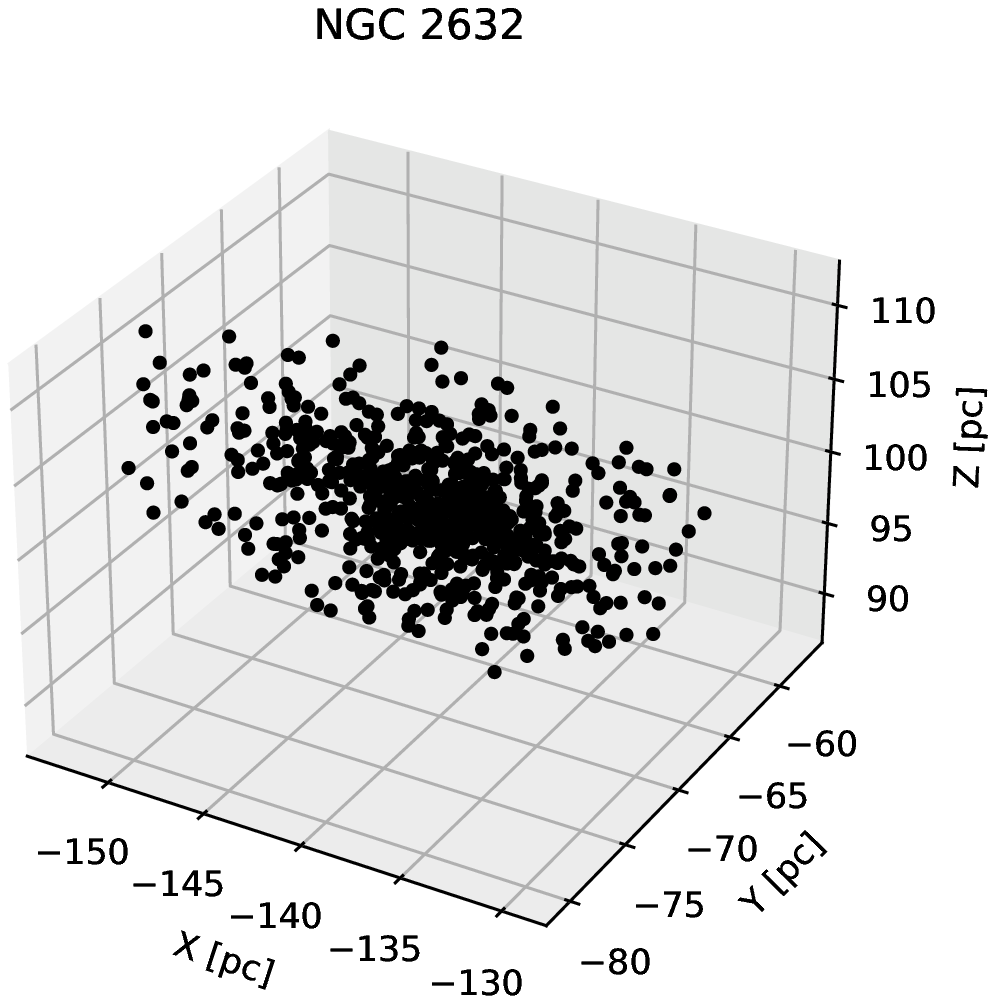}} \quad & 
   \subfloat[][NGC 5823 at $d$\,=\,1813(11)\,pc. \label{NGC_5823}]{\includegraphics[scale=0.48]{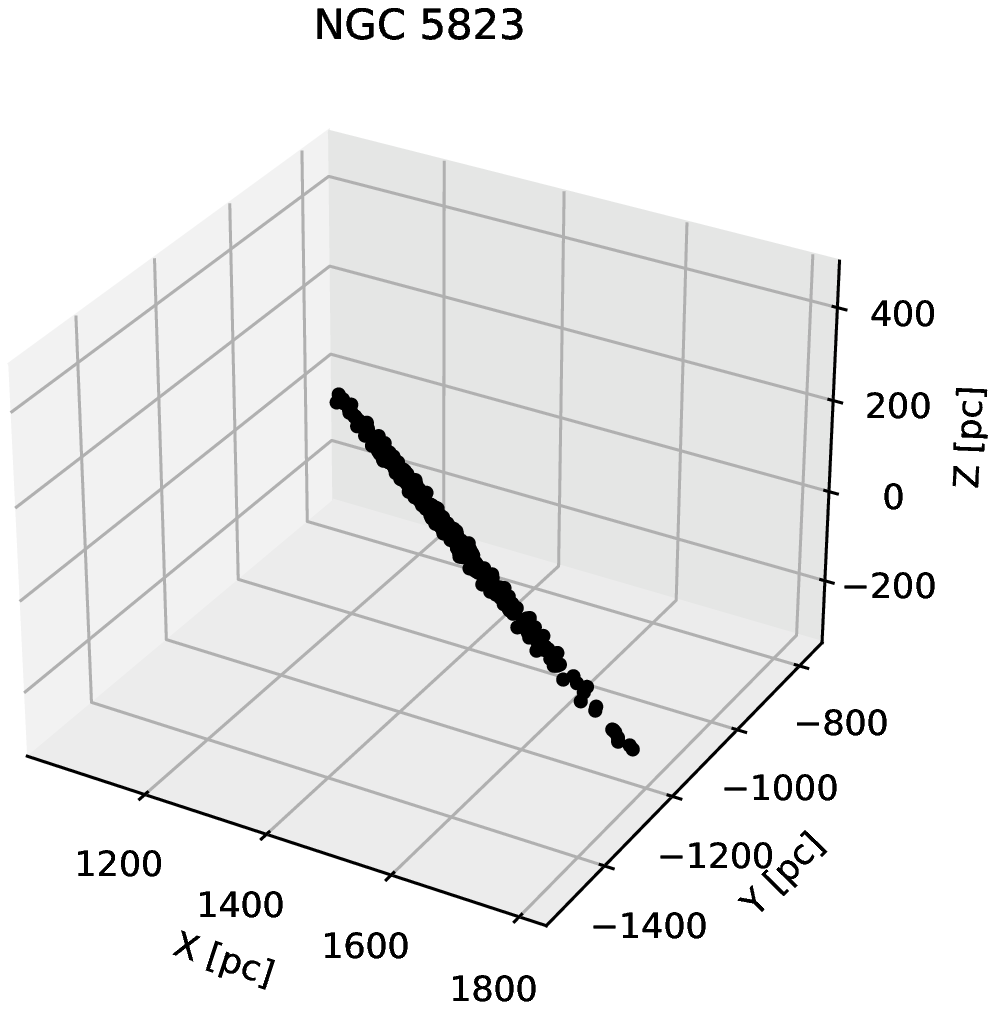}} \\
	 \end{tabular}
   \caption{The 3D structures of the clusters indicated. The expected needle-like structure in the line-of-sight is clearly visible for more distant clusters.}
   \label{3D_four_clusters}
\end{figure*}

We should keep in mind that the size of the cluster determined from projected widths depends heavily on the
procedure that was used to determine the cluster membership probabilities. A good example of this can be seen if we compare our results for
NGC 2682 with another work which focuses on studying this particular cluster. While our cluster width turns out to be
about 3 pc, results from Carrera et al. (2019) show that the
size of the cluster is about one order of magnitude larger. This can be expected since it is known that this cluster is
very old ($\log{t} \sim 9.5$, Bossini et al. 2019) and has experienced a significant amount of dynamical evaporation (Carrera et al. 2019).
The radius of a cluster should increase with time. We conclude that our calculated widths of clusters contain
some systematic errors which should be negligible for the youngest clusters and get significantly larger for much older clusters.
We expect that this would affect the distribution in Fig.~\ref{widths_OCLs} by slightly enhancing the size of the tail toward larger
projected widths at the expense of lowering the peak at lower values of the distribution.

Finally, we have also calculated the values of the sigma parameter $\sigma_r$ from $\sigma_\varpi$ using both, inverse-parallax
approach as well as the distances from Bailer-Jones et al. (2018), who included two parts into the prior of their analysis -- the exponentially
decreasing volume density term and a Galactic model term. Although their approach is not best suited for studying open clusters, it still gives us
a different look at the sigma parameter (especially for the comparison with the very different inverse-parallax approach). In Fig.~\ref{distance_sigma1},
we present the dependency of the sigma parameter $\sigma_r$ on the distance $r$ for the loose
sample (upper panel), derived using the Bailer-Jones distances $r_{\mathrm{B-J}}$ (the cluster distance is calculated as median of $r_{\mathrm{B-J}}$).
The most distant cluster is Teutsch~106 with a distance of about 6\,kpc from the Sun.
A closer inspection yields that the data up to 2\,kpc (lower panel of Fig.~\ref{distance_sigma1}) allow to study
the outliers (in those plots) in more details. A quadratic fit of the sigma parameter (SGP) in the closer inspection yields

\begin{equation}
\textrm{SGP}_{\mathrm{B-J}} = 1.48(46) + 0.011(2) \, r_{\mathrm{B-J}} + 0.0000400(18) \, r_{\mathrm{B-J}}^2 \,\,,
\end{equation}
with a standard error of 14.9\,pc, respectively. This transforms to a SGP of [3, 7, 17, 53, 184\,pc] for distances of
[100, 250, 500, 1000, 2000\,pc] not taking into account the derived standard deviation. In total, we found 50 open clusters which
exceed 3$\sigma$ above the standard line. These aggregates are good candidates for either hosting two populations in the same
line-of-sight or not being true star clusters. Within 1\,kpc, we find five of them: Alessi 44 (ASCC 106; most deviating case), NGC 1579, 
NGC 2183, NGC 6178, and vdBergh 80. For these open clusters,
we find no conspicuous features. However, these 50 aggregates have to be investigated in more details using photometric data
and the available results from the literature to shed more light on the inconsistencies.

On the other hand, we find several clusters which are 3$\sigma$ below the standard line and are therefore very well defined.
In principle, these clusters could be the best candidates for studying the individual three-dimensional structures. However,
it is advisable to first analyse the colour-magnitude diagrams which should help to lower the field-star contamination.

When using the inverse-parallax approach, the situation does not significantly change (Fig.~\ref{distance_sigma2}). We find the quadratic fit of SGP (again, for the plot in the closer inspection)

\begin{equation}
\textrm{SGP}_{\mathrm{inv}} = 1.94(56) + 0.009(3) \, r_{\mathrm{inv}} + 0.0000429(22) \, r_{\mathrm{inv}}^2 \,\,,
\end{equation}
with a standard error of 15.3\,pc, and SGP of [3, 7, 17, 54, 192\,pc] for distances of
[100, 250, 500, 1000, 2000\,pc]. These results are very similar to the case when we used the Bailer-Jones distances.
Although the numerical results are somewhat different, the elongation of the clusters (measured by SGP) does not significantly
differ from the previous method used to derive the distances. The values of $\textrm{SGP}_{\mathrm{inv}}$ start to notably deviate from
$\textrm{SGP}_{\mathrm{B-J}}$ only at distances starting from about 2.0~kpc and beyond.

\subsection*{3.2. Characterizing clusters in three spatial dimensions} \label{subsection_3D}

The three dimensional spatial structure of open clusters based on observations is very much needed
for all cluster formation and evolution models (Kroupa 1995). Open questions, like the internal kinematical and spatial distributions 
of the members and their evolution, can only be answered by detailed observations of open clusters of different 
ages. Up to now, there are only very few of such investigations on the basis of \textit{Gaia} DR2 data available 
(Franciosini et al. 2018, Karnath et al. 2019). This motivated us investigating the 3D characteristics on the basis
of the currently available data and their errors. Particularly we are interested up to which distances such
an analysis is meaningful.

For a given cluster, the coordinates $d_1$ (or $r$) and $d_2$ of its individual members form a distribution which can be displayed in histograms. 
These can give us insights about the spatial structure of the cluster. As we have seen, if a given cluster is located within 2.0~kpc, the SGP does
not depend too much on the method we use for deriving the distances. Although such approach does not allow us to study clusters in detail at large distances,
we can still say something about the effect of the SGP on the 3D structure (from a statistical point of view).

\begin{figure}[t]
\centering
\includegraphics[scale=0.5]{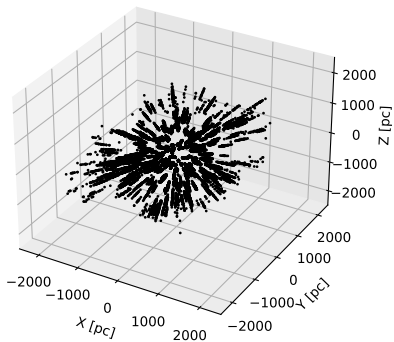}
\caption{The shape of the star clusters from the loose sample in the Galactic [$X\,Y\,Z$] coordinate system.}
\label{3D_strict_sample} 
\end{figure}

\begin{figure}[t]
\centering
\includegraphics[scale=0.5]{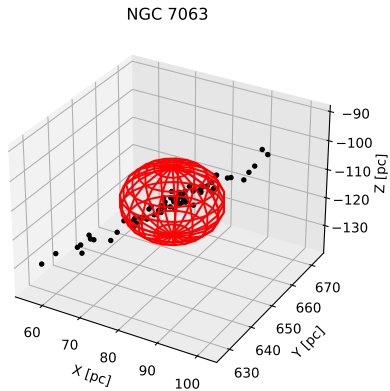}
\caption{The 3D structure of NGC 7063 at $d$\,=\,665(2)\,pc for all stars with a membership probability larger
than 50\%. The elongation is the result of uncertainties in the observed parallaxes. Included is a sphere with a radius of 12\,pc centered at the
star cluster.}
\label{NGC_7063} 
\end{figure}   

Using the data from \textit{Gaia} DR2 and the distances $r_{\mathrm{B-J}}$, we are able to create scatter plots using the [$X\,Y\,Z$] coordinates of cluster members. For distant 
clusters, we would expect the 3D structure to be needle-like, because of the absolute values of the errors in distances. It is 
interesting to see, that we can see this structure quite clearly even at distances $<$750\,pc. Nearby highly-populated clusters 
(like NGC 2632) with $r$\,$<$\,250\,pc, on the other hand, show only a very weak elongation in the line-of-sight. This tells us something 
about how even the best data (we currently have) limit the investigation of 3D structures of open clusters. In Fig.~\ref{3D_four_clusters},
we show the situation for four clusters: NGC 1039 (distance of 505(2)\,pc), NGC 1528 (1021(4)\,pc) NGC 2632 (186($<1$)\,pc), and NGC
5823 (1813(11)\,pc). The expected needle-like structure in the line-of-sight is clearly visible. One has to keep
in mind, that the apparent members of NGC 5823 are spread about 160\,pc around the Galactic disk, for example. To investigate this
topic further, we have plotted all star clusters from the strict sample in the Galactic [$X\,Y\,Z$] coordinate system (Fig.~\ref{3D_strict_sample}) which
corresponds to Fig. 2 shown in Ward et al. (2020).
The described limitations might be one of the reason why we are not able to precisely trace the spiral arms with open clusters 
(see Fig.~11 in Cantat-Gaudin et al. 2018).  

\begin{figure}[t]
\centering
\includegraphics[scale=0.5]{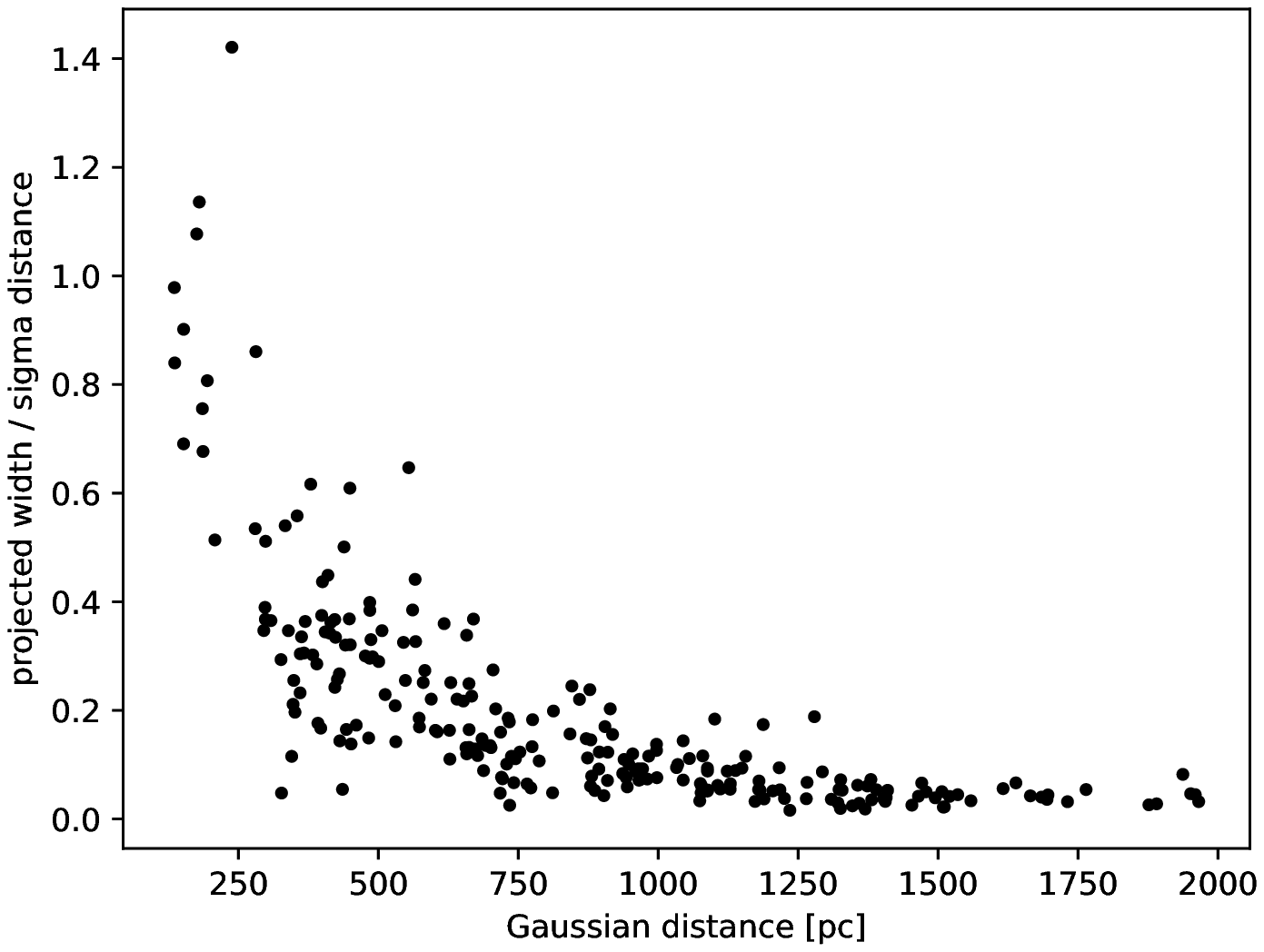}
\caption{The plot of a measure of ellipticity of clusters against the distance of the clusters (strict sample).}
\label{eccentric} 
\end{figure}

We would like to show the impact of those results (using the loose sample) on the estimates of 3D radii of the individual clusters. We have 
fitted each cluster with a sphere which contains 50\% of the population of a given cluster (the centre of the cluster is taken to be 
the median value of the [$X\,Y\,Z$] coordinates). These spheres can be then compared with the reference sphere of radius 12\,pc and we can 
search for the sphere which is smaller than (or approximately equal to) 12\,pc and is the most distant. In Fig.~\ref{NGC_7063} we show the case of NGC 7063 with a 
distance of 665(2)\,pc from the Sun. All stars with a membership probability larger than 50\,\% are included in this figure.
Again, the distribution of the members is not spherical but more needle-like.
This should serve as an upper limit to the distance up to which we can still somewhat fit spheres to open clusters.
However, it has to be emphasized that the true internal 3D structure of such an open cluster cannot be studied. This is mainly due to the elongation
resulting from the observational uncertainties. However, there is also a secondary effect that has to be taken into account -- the bias
introduced by the search of members in the line-of-sight. In the future, algorithm searching for members of more distant star clusters
should transform the astrometrical data in the [$X\,Y\,Z$] space and not using direct line-of-sight distances. This would guarantee to search
for members in a three-dimensional space around the cluster centre and overcome possible (although small) selection effects.  

In this respect we have to think about the 
definition of a star cluster and the differences to a moving group. Recently, Faherty et al. (2018) presented an analysis of
a co-moving catalogue including 4555 groups of stars (10606 individual objects). Questions arise like what is the lowest
number of members and the lowest total mass of a star cluster, for example. And how can we distinguish between a moving group
and a star cluster?  It seems that the distribution of the stars (central agglomeration) is only a poor criterion to do so.
Here, new comprehensive methods are needed.

It is expected that star clusters should have a finite lifetime resulting from the dynamical evaporation process (Chumak et al. 2010).

\begin{figure*}
  \centering
  \includegraphics[scale=0.5]{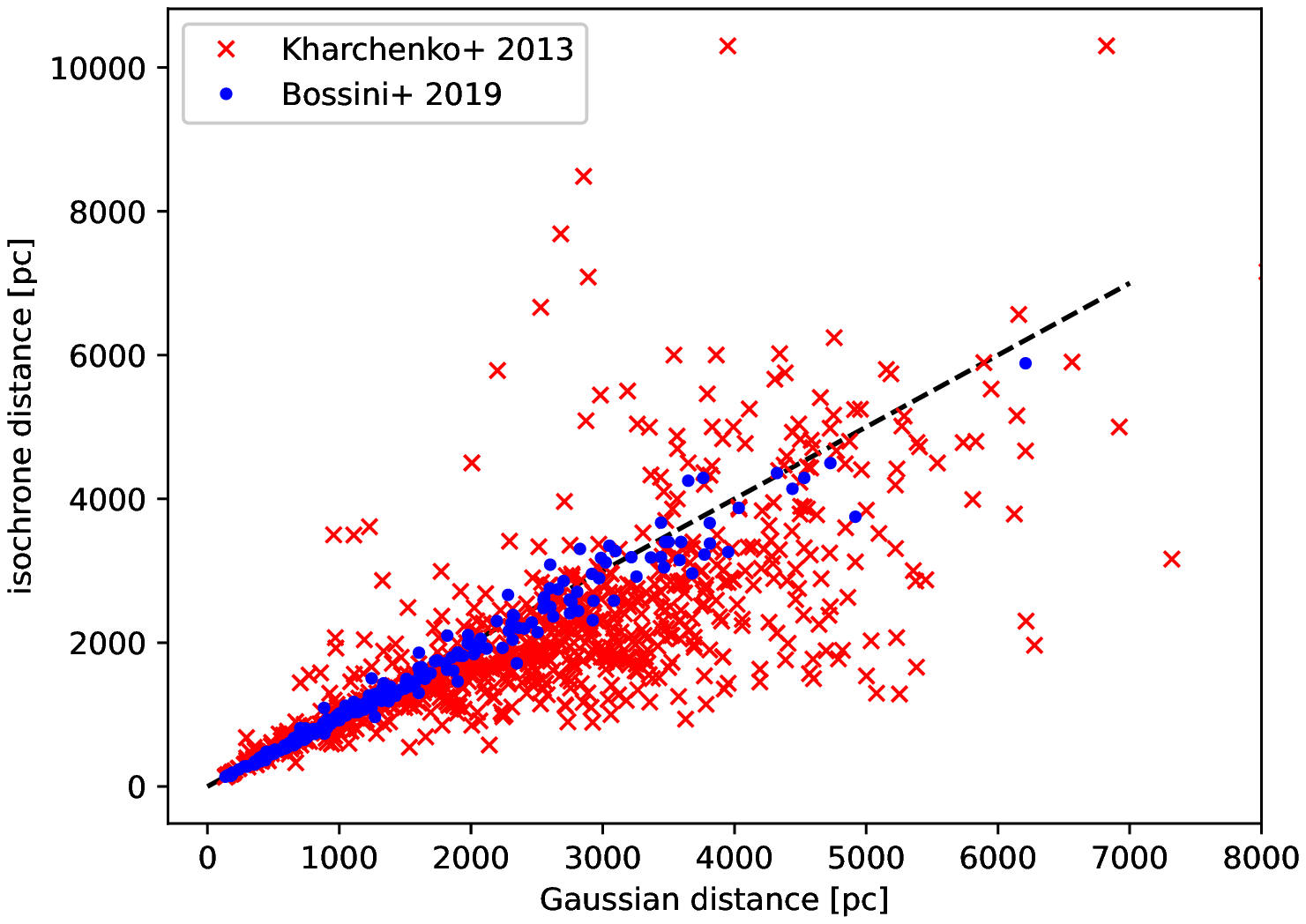}
  \includegraphics[scale=0.5]{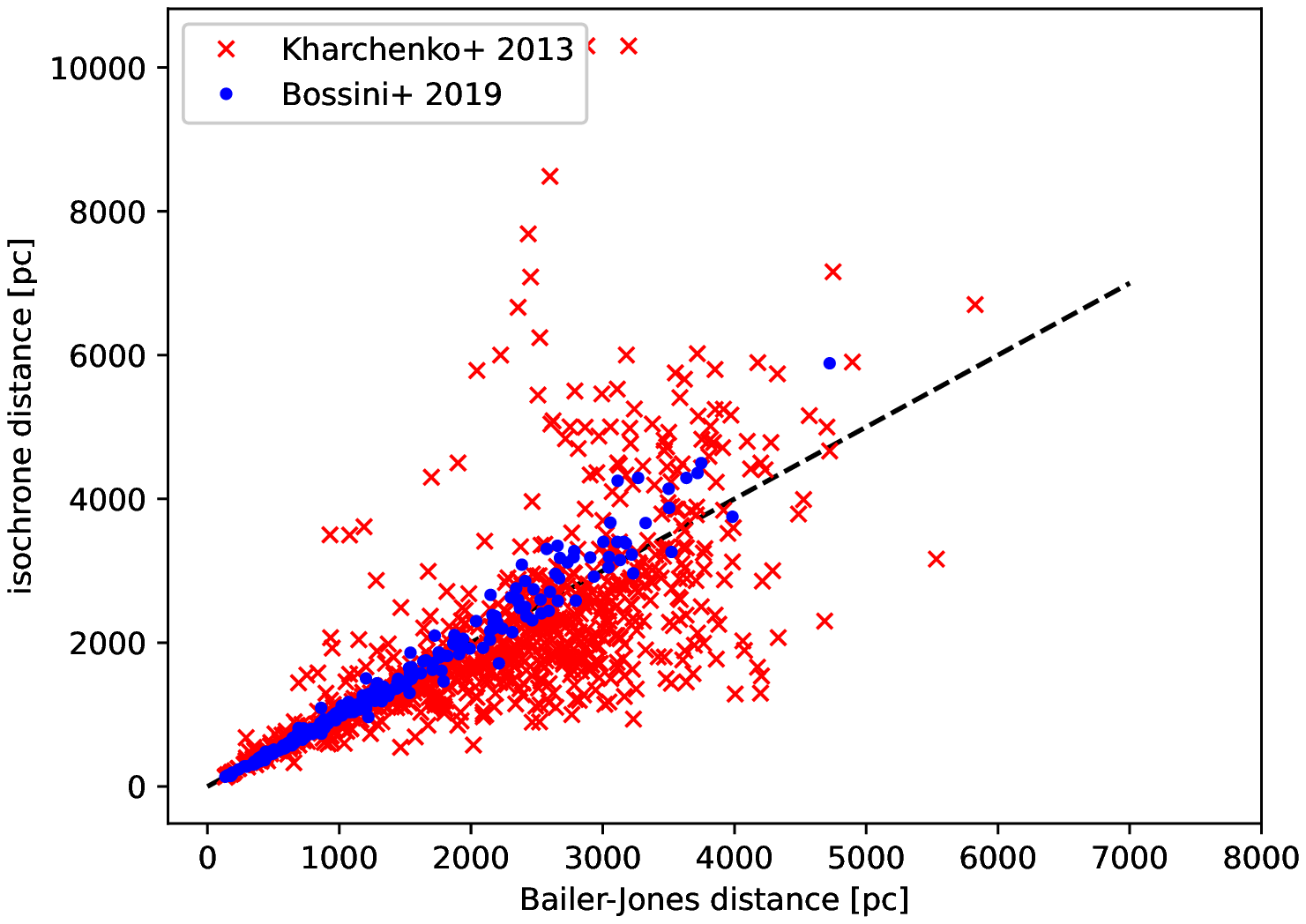}
  \caption{Comparison of the cluster distances derived using two different isochrone fitting techniques with the distances calculated from the inverse values of the central locations of the parallax distributions (upper panel) and from the median of the Bailer-Jones distances (lower panel). The black dashed line represent one-to-one correlaton.}
  \label{iso_BailerInverse}
\end{figure*}

Another well known effect are tidal tails of star clusters. It is expected that star clusters should have a finite lifetime resulting
from the dynamical evaporation process (Chumak et al. 2010). Especially for old open clusters, the
total mass at birth is difficult to establish because of the member loss over several hundreds of millions of years. But estimating
this parameter is important for putting constraints on the star forming rate in the Milky Way. Recently, tidal tails were investigated
for the close open clusters Gamma Velorum (Franciosini et al. 2018), Hyades (R{\"o}ser et al. 2019), and Praesepe (R{\"o}ser \& Schilbach 2019), 
for example. The search for such tidal tails is not straightforward because according to models they could reach up to
a length of about 800\,pc, as in the case of the Hyades (Ernst et al. 2011). To find these tails, one has to use a 
method calculating the space velocities to a convergent point. However, because measured radial velocities are missing for the vast
majority of stars, the before mentioned works rely on criteria solely based on tangential velocities. If we investigate our sample,
in the case of most clusters the distance errors are far too high for performing such an analysis. A good example is M~67 (distance of
about 800\,pc from the Sun), for which
Carrera et al. (2019) reported an extended halo of up to 150\,pc which transforms to 10$^{\deg}$ on the sky. Such large extended searches
for cluster members meet the limits for any automatic method. We also have to emphasize that the needle-like structures seen in
Figs. \ref{3D_four_clusters} and \ref{3D_strict_sample} are not due to tidal tails but due to the uncertainties in observed parallaxes
of the individual members of star clusters.

In order to further analyse the limitations of the data for studying clusters we may need to find a measure of the ellipticity of clusters. For this, we have chosen to take the ratio of projected widths and distance sigma parameters (based on $r_{\mathrm{inv}}$). In Fig.~\ref{eccentric} we see how this measure behaves as a function of the distance. We have used the strict sample to get the best possible results. It can be clearly seen that the clusters from our sample significantly differ from spherical symmetry already at 500 pc.

\subsection*{3.3. Comparison with isochrone fitting techniques}

In order to fully understand the quality of \textit{Gaia} DR2 astrometric data, we would like to compare our results with the distances from literature that were derived using isochrone fitting techniques. Isochrones present us an option of comparing distances calculated from two independent methods. For this purpose, we have taken the data from Kharchenko et al. (2013) and Bossini et al. (2019) and compared their distances with the distances we found using $r_\mathrm{inv}$ and $r_\mathrm{B-J}$.

In the upper panel of Fig.~\ref{iso_BailerInverse} we see that the cluster distances determined from the inverse-parallax approach very well correlate with the distances from isochrone fitting. However, there are some apparent differences. First of all, there is an apparent offset between the distance values from Bossini et al. (2019) and those we calculated. When compared with Kharchenko et al. (2013), our distances seem to be somewhat over-estimated, which is especially clear at distances larger than 2.0 kpc.

The lower panel of Fig.~\ref{iso_BailerInverse} shows us that the Bailer-Jones distances are better correlated with those from Kharchenko et al. (2013) than in the previous case. The offset is gone when plotted against the data from Bossini et al. (2019), but the correlation at larger distances appears to be worse -- here $r_{\mathrm{B-J}}$ seems to be somewhat underestimated.

\begin{figure*}
  \centering
  \includegraphics[scale=0.5]{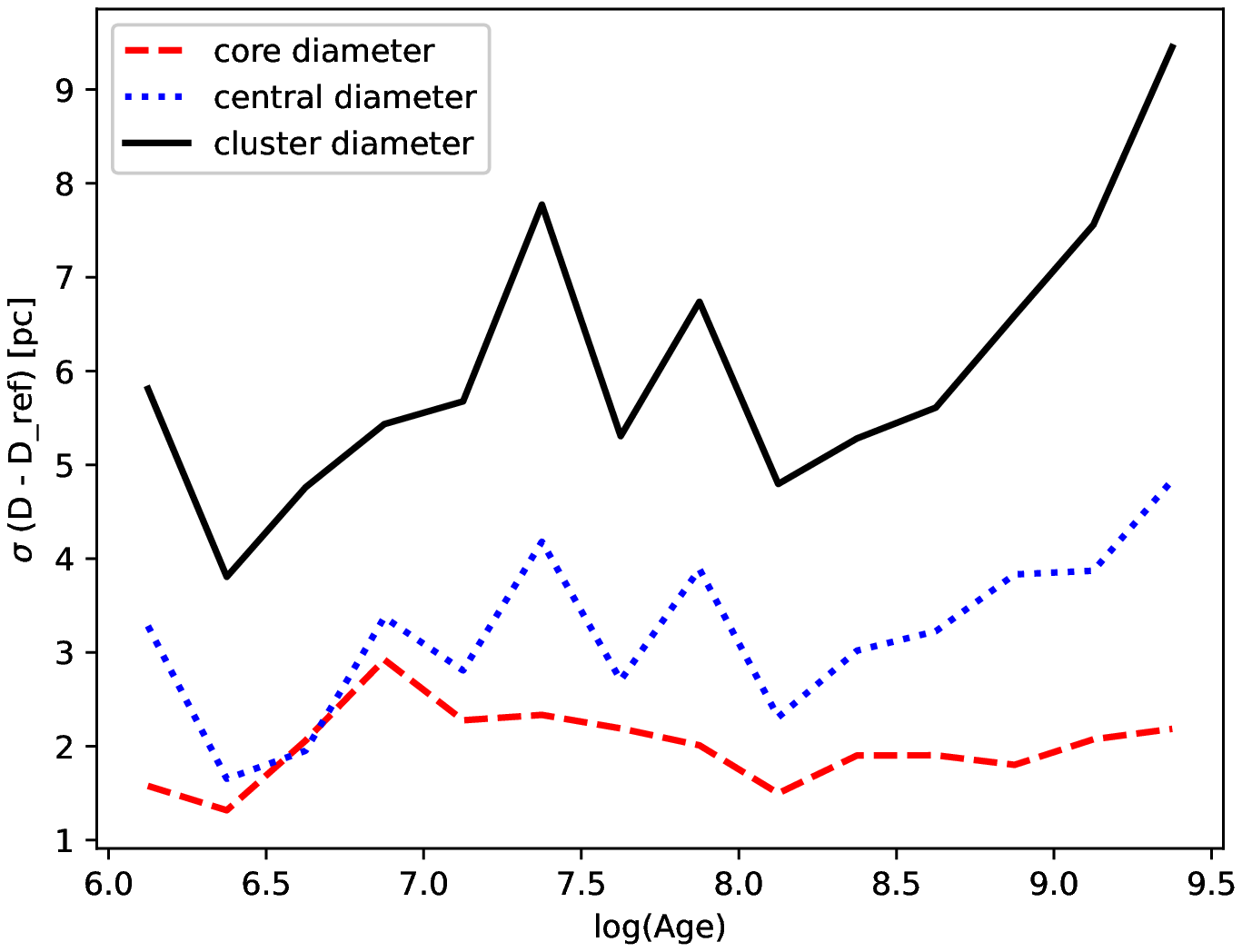}
  \includegraphics[scale=0.5]{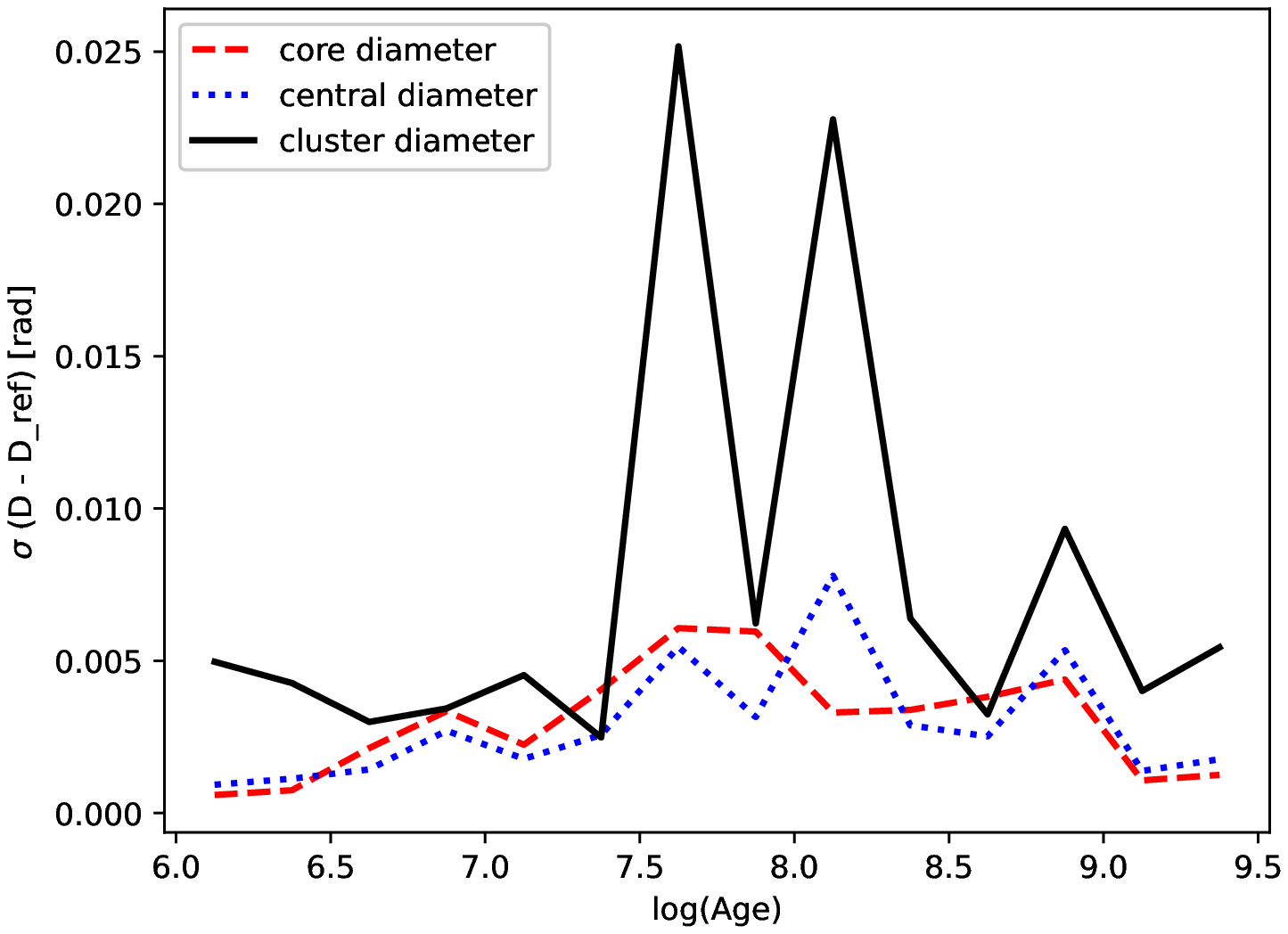}
  \caption{Scatter of the differences between diameters determined in this work and those from Kharchenko et al. (2013). The scatter is measured by the standard deviation of diameter differences at a given range of logarithmic ages (step $\Delta \log{(\textrm{Age})}=0.25$). The upper panel shows absolute diameters. In the lower panel, angular diameters are displayed.}
  \label{iso_DiametersSigma}
\end{figure*}

Finally, we would like to verify our suspicion that the values of cluster diameters (discussed above) are underestimated. The main reason for our assumption is the dissipation of clusters -- for older clusters, we would expect much higher values of diameters. We can check this by looking at the diameters and ages derived by Kharchenko et al. (2013). The problem is that we cannot simply look at the differences of diameters since the definitions of the cluster radii (and diameters) in Kharchenko et al. (2013) differ from the approach we used in this work (projected widths, discussed in previous subsections). Instead, we want to see how the standard deviations of the diameter differences at a given range of ages depend on the logarithmic age. We have plotted this relationship (Fig.~\ref{iso_DiametersSigma}, upper panel) for all three radii defined in Kharchenko et al. (2013). This result seems to confirm our suspicion -- it seems that the cluster members taken from see Fig.~11 in represent only the core population of the studied clusters. However, the disagreement between the distances (Fig. \ref{iso_BailerInverse}, upper panel) is going to affect this result. In Fig.~\ref{iso_DiametersSigma} (lower panel), we have plotted the angular diameters of the cluster in the same way as before. In this case, there is no clear scatter at $\log{(\textrm{Age})}>8.5$.

\subsection*{3.4. Simulating SGP}

We predict that the shape of the function SGP($r$) is determined by the parallax uncertainty. This can be easily verified by simply simulating a number of clusters at a random distance. We have chosen to simulated 500 clusters (containing between 100 and 300 members) at distances between 100 pc and 2000 pc, which corresponds to the region shown in lower panels of Fig.~\ref{distance_sigma1} and Fig.~\ref{distance_sigma2}. The cluster radii were chosen to be 5 pc.

When the clusters are created ("real" parallaxes are found by inverting the simulated "real" distances), "observed" parallaxes are simulated using a random (normal) function based on two possibilities -- either the absolute or the relative error is set to be a constant. Afterwards, the distances and SGP values are determined by fitting Gaussian functions, as described previously.

\begin{figure}
  \centering
  \includegraphics[scale=0.5]{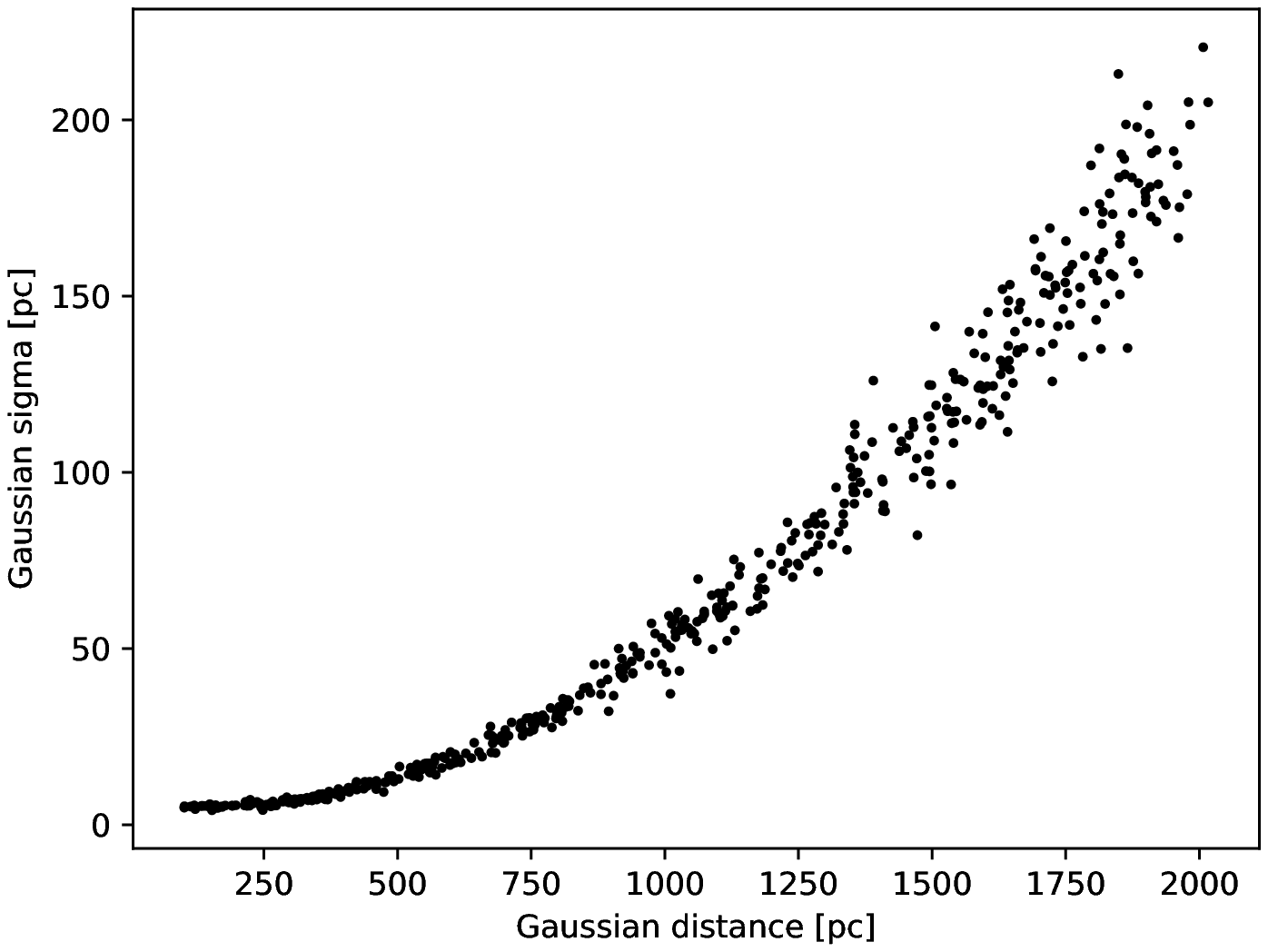}
  \includegraphics[scale=0.5]{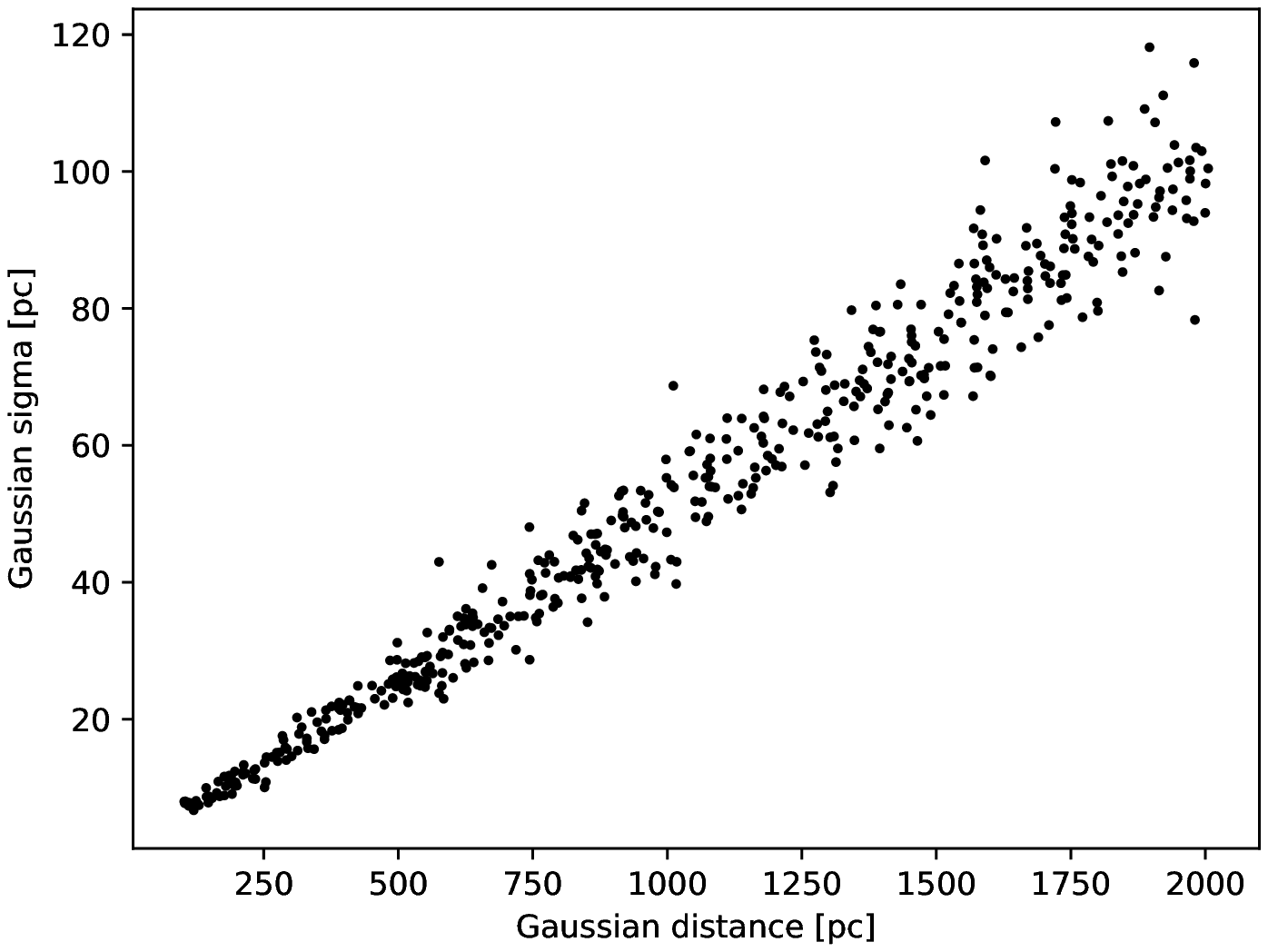}
  \caption{The relation between the SGP and the distance of a cluster based on a simulation. The upper panel shows the example of a constant absolute parallax error ($\sigma_{\varpi}=0.05$ mas). The lower panel assumes a constant relative parallax error ($\sigma_{\varpi} / \varpi = 0.05$).}
  \label{simulatedSGP}
\end{figure}

The results for $\sigma_{\varpi}=0.05$ mas and $\sigma_{\varpi} / \varpi = 0.05$ are displayed in Fig.~\ref{simulatedSGP}. The first situation coincides (almost exactly) with the results of our analysis of the clusters from Cantat-Gaudin et al. (2018) -- the quadratic relation is clearly the result of the absolute parallax uncertainties. The larger spread in the observed data, when compared with the simulated data, is most likely the result of variations in the value $\sigma_{\varpi}$ (an estimate of the distribution of these values is shown in Fig.~\ref{fignew3}).

\section*{Conclusions} \label{conclusions}
 
With the most recent \textit{Gaia} DR2, it is now possible to study nearby open clusters in more details.
Especially the internal structure and kinematical characteristics are still only known for a very
few clusters like the Hyades and Pleiades. But these characteristics are very important as input parameters for models
dealing with the formation and evolution of star clusters.

Cantat-Gaudin et al. (2018) studied 1229 open clusters and derived membership probabilities of stars as well as cluster
distances and diameters based on astrometrical and kinematical data. We used the cluster members from these data to study the
limitations of the \textit{Gaia} DR2 when it comes to study open clusters. The distances in this work were determined by using the most
typical procedures -- the inversion of parallaxes and the Bayesian method with decreasing volume density prior. For the second procedure, we used
the values presented by Bailer-Jones et al. (2018) who also included a Galactic model in their prior which influences the distances of the open clusters
by slightly underestimating these values. The comparison of
the two data sets of distances with the isochrone fitting methods shows that the calculated distances are in a good agreement. Together
with the simulations of clusters presented in Sect.~\ref{r_to_plx}, this shows that the distances used in this work are quite reliable, at least in a statistical
sense.

Due to the uncertainties in observed parallaxes, most of the clusters have 
needle-like shapes and are not even close to being spherical, which can be expected when comparing with the results from
Luri et al. (2018). We conclude that this affects the determination of distances not only
when using inverted parallaxes but also when the Bayesian approach with decreasing volume density prior
is applied. The use of the Galactic model in the prior in Bailer-Jones et al. (2018) seems to have
little to no effect on the apparent elongation of clusters along the line of sight. It is possible that the situation will improve when a better
prior is used, like the one mentioned in Carrera et al. (2019).

With the current available data, the diameters of open clusters can be well studied up to about 2\,kpc (using a 
statistical approach). The results of the overall distribution are in line with the current models showing that all 
clusters have diameters less than 20\,pc with a peak value lying between 2 and 4 pc. However, this
result depends critically on the method used to determine the cluster membership probabilities. Furthermore,
we find that individual open clusters beyond 500\,pc should not be considered for 3D studies with the most widely
used parallax-to-distance transformation methods.

Comparison of the derived cluster distances with isochrone fitting methods shows that both approaches give
statistically very similar results (except when we try to compare distances from \textit{Gaia} with isochrones derived
from older data sets). Looking at the comparison of the derived projected widths with the diameters from Kharchenko et al. (2013),
we find no evidence that would show an expected systematic increase of the cluster diameters with the increasing cluster ages.
The most likely explanation is that both sets of cluster members, those from Kharchenko et al. (2013) and Cantat-Gaudin et al. (2018),
fail to include the outermost members.

The work by Cantat-Gaudin \& Anders (2020) provided additional clusters when compared to Cantat-Gaudin et al. (2018). However, the previous clusters remain unchanged. For this reason we argue that the inclusion of the updated data set should not significantly change the statistical results of this analysis. On the other hand, the data for the individual stars in clusters have slightly changed in the recently released \textit{Gaia} EDR3 (Gaia Collaboration 2020). From the statistical point of view, we do not expect anything to change, although this prediction has to be verified once the new set of clusters (based on the new data) has been released.

With the new data sets (e.g. \textit{Gaia} DR2 or EDR3), the definitions of an open cluster and of a moving group have to be revised.
Quantities like the lower limit of the number of cluster members and total masses have to be assessed anew.

\section*{Acknowledgements}

This work has been supported by MUNI/A/1482/2019 and MUNI/A/1206/2020 (Masaryk University, Faculty of science), and the Erasmus+ programme of the 
European Union under grant number 2020-1-CZ01-KA203-078200.
This research has made use of the WEBDA database, operated at the Department of Theoretical Physics and Astrophysics of the Masaryk University, 
the SIMBAD database, operated at CDS, Strasbourg, France and NASA's Astrophysics Data System.
This work presents results from the European Space Agency (ESA) space mission Gaia. Gaia data are being processed by the Gaia Data Processing and 
Analysis Consortium (DPAC). Funding for the DPAC is provided by national institutions, in particular the institutions participating in the Gaia 
MultiLateral Agreement (MLA). The Gaia mission website is https://www.cosmos.esa.int/gaia. The Gaia archive website is https://archives.esac.esa.int/gaia.

%\newpage

\end{document}